\documentclass[%
superscriptaddress,
 amsmath,amssymb,
 aps,
prl,
twocolumn
]{revtex4-2}

\usepackage{graphicx}% Include figure files
\usepackage{dcolumn}% Align table columns on decimal point
\usepackage{bm}% bold math
\usepackage{xcolor,soul}

\usepackage{hyperref}% add hypertext capabilities
\usepackage[normalem]{ulem}

\DeclareRobustCommand{\vect}[1]{
  \ifcat#1\relax
    \boldsymbol{#1}
  \else
    \mathbf{#1}
  \fi}

\usepackage{graphicx} % Required for inserting images
\usepackage{tabularx}
\usepackage{mathtools}
\usepackage{amsmath}
\usepackage{amssymb}
\usepackage{comment}
\usepackage{bm}
\usepackage{graphicx}
\usepackage{xcolor}
\usepackage{epstopdf}
\usepackage{dcolumn}
\usepackage{bm}
\usepackage{amsfonts}
\usepackage{array}
\usepackage{amsbsy}
\usepackage{color}
\usepackage{hyperref} 
\usepackage{enumerate}
\usepackage{lipsum} 
\usepackage{bigints}
\usepackage{comment}

  \newcommand{\av}[1]{\left\langle#1\right\rangle}
  \newcommand{\cbr}[1]{\left(#1\right)}
  \newcommand{\sbr}[1]{\left[#1\right]}

\newcommand{\op}{\varepsilon}
\newcommand{\cP} {\mathcal{P}}
\newcommand{\lx} {\left}
\newcommand{\rx} {\right}

\begin{document}

%\title{Reversible stochastic processes in the absence of parity rules}
\title{Thermal equilibrium with generalized time-reversal symmetry}
%\title{Langevin dynamics with generalized time-reversal symmetry}
\author{Dario Lucente}
\affiliation{Department of Mathematics \& Physics, University of Campania “Luigi Vanvitelli”, Viale Lincoln 5, 81100 Caserta, Italy}
\author{Marco Baldovin} \email{marco.baldovin@cnr.it}
\affiliation{Institute for Complex Systems, CNR, Universit\`a Sapienza, I-00185, Rome, Italy}
\author{Massimiliano Viale}
\affiliation{Institute for Complex Systems, CNR, Universit\`a Sapienza, I-00185, Rome, Italy}
\author{Angelo Vulpiani}
\affiliation{Institute for Complex Systems, CNR, Universit\`a Sapienza, I-00185, Rome, Italy}
\affiliation{Department of Physics, University of Rome “La Sapienza”, P.le Aldo Moro 5, 00185 Rome, Italy}
\date{\today}

\begin{abstract}
In the study of stochastic processes, identifying the parity under time-reversal is essential to verify detailed balance, and to compute the entropy production rate (which is, otherwise, ambiguously defined). While in many cases the correct time-reversal symmetry is suggested by physical arguments, for generic processes the identification is not trivial: as a result, systems at thermal equilibrium may be mistakenly interpreted as non-equilibrium ones. 

We focus on the reversible deterministic dynamics of a slow variable coupled to many degrees of freedom acting as a thermal bath. We show that the time-reversal symmetry of the slow variable is preserved when passing to an effective stochastic description, independently of the nature of the bath. In turn, for generic 2-dimensional continuous Markov processes, we provide a criterion to identify the time-reversal parity rules under which the dynamics is at equilibrium (if any).
The case of the Lotka-Volterra model is discussed as an example.

\end{abstract}

\maketitle

\textit{Introduction ---} 
Parity under time reversal (TR) plays a fundamental role in the description of nonequilibrium phenomena~\cite{dieball2025perspective}. TR operators explicitly appear in the definition of detailed balance and entropy production~\cite{lebowitz1999gallavotti,maes2000definition}, as well as in the derivation of Fluctuation Relations~\cite{gallavotti1995dynamical,kurchan1998fluctuation,maes2003origin,speck2005integral}.
For stochastic dynamics, TR symmetry is not identified from the evolution equations: it originates instead from their physical interpretation. This is a difference with deterministic systems, where reversibility is unambiguously defined by the existence of a reversing symmetry of the flow~\cite{lamb1998time}. 
Consider as an example the linear stochastic differential equations
\begin{equation}
\label{eq:linear}
\begin{cases}
\dot{x}_1=- x_1+x_2 +\xi_1 \,\\
\dot{x}_2=-x_1- x_2+\xi_2\,,
\end{cases}
\end{equation}
where $\av{\xi_{i}}=0$, $\av{\xi_{i}(t)\xi_{j}(t')}=2\delta_{ij}\delta(t-t')$. If $x_1$ and $x_2$ are interpreted as (dimensionless) position and momentum, with opposite TR parity, then~\eqref{eq:linear} represents an underdamped harmonic oscillator and the system is at equilibrium. If, instead, both $x_1$ and $x_2$ are even, then Eq.~\eqref{eq:linear} models a pair of overdamped particles with non-reciprocal interactions (or, equivalently, a Brownian Gyrator~\cite{lucente2022inference,lucente2024conceptual}), and the dynamics is out of equilibrium~\cite{loos2020irreversibility}. 

Stochastic processes may originate from a wide variety of physical situations~\cite{krapivsky}, and TR rules will in general depend on the considered phenomenon.
Van Kampen identifies, in particular, two ways of introducing  stochastic processes~\cite{10.1007/BFb0105593}: via large-deviations theory methods for macroscopic observables, or from deterministic dynamical systems (multiscale expansions and homogenization techniques)~\cite{zwanzig2001nonequilibrium,Cpas1998,guioth2022path}. A consistent notion of TR which follows the former prescription is defined within Macroscopic Fluctuation Theory~\cite{bertini2015macroscopic,falasco2025macroscopic}. In this Letter we focus instead on the latter, i.e. on stochastic processes modeling the effective evolution of slow degrees of freedom in multiscale deterministic systems at equilibrium~\cite{pavliotis2008multiscale, zwanzig2001nonequilibrium,Cpas1998,guioth2022path}. More precisely, we consider a slow deterministic dynamics coupled with many fast particles, acting as a thermal bath.  By restricting to TR symmetries that only act on the variables, without affecting external fields~\cite{bonella2015time,bonella2017time}, we  show that if the global, deterministic system is reversible, then the slow stochastic process is also reversible. The symmetry is the same as in the deterministic limit (i.e., when the slow particle is uncoupled from the bath).

This first result allows us to formulate a criterion, valid for 2-dimensional (2D) Langevin processes, to establish if a stochastic dynamics is reversible in the above sense: i.e., if it can be seen as the effective evolution of a slow (observable) degree of freedom, at equilibrium with a fast (hidden) thermal bath. This criterion can be used to assess or exclude the presence of equilibrium even if the TR symmetry is not \textit{a priori} known.
While equilibrium conditions for stochastic processes under generic TR operators have been studied before~\cite{chetrite2008fluctuation,chetrite2008fluctuation2,gawedzki2013fluctuation,o2024geometric,o2024geometric2}, this method is, to the best of our knowledge, a new result.

\textit{Conditions for reversibility--- } 
The dynamical system 
\begin{equation}
\label{eq:dyn}
\frac{d \vect{x}}{d t} = f(\vect{x})\,
\end{equation}
is \textit{reversible}~\cite{lamb1998time} if  a volume-preserving transformation $\widehat{\vect{x}}={\op}( \vect{x})$ exists, such that ${\op}( \widehat{\vect{x}})=\vect{x}$ (involution property) and
\begin{equation}
\label{eq:reversing_symmetry}
\frac{d \widehat{\vect{x}}}{d t} =- f ( \widehat{\vect{x}})\,.  
\end{equation}
We will focus on systems where a quantity  $H(\vect{x})$, with the physical meaning of an energy, is conserved by the dynamics, and $H(\widehat{\vect{x}})=H(\vect{x})$.
For any reversible system~\eqref{eq:dyn}, there is a class of continuous stochastic processes that admit it as the deterministic limit. They obey stochastic differential equations of the form
\begin{equation}
    \label{eq:lan}
    \dot{\vect{x}} =f(\vect{x})+\gamma\Gamma(\vect{x})+\vect{\xi}(t)\,,
\end{equation}
in the It\={o} covention, where $\Gamma$ is a function of the phase space, $\vect{\xi}$ is a Gaussian white noise such that 
\begin{equation}
\label{eq:noisecorr}
\av{\boldsymbol{\xi}}=0\,,\quad \av{\boldsymbol{\xi}(t)\boldsymbol{\xi}^T(t')}=2\gamma D(\vect{x})\delta(t-t')\,,
\end{equation}
$D$ is a symmetric $n\times n$ matrix, and $\gamma$ is a constant. The deterministic  limit~\eqref{eq:dyn} is recovered for $\gamma\to0$. In many physical situations, the case $\gamma > 0$ accounts for the presence of an external bath, perturbing the isolated dynamics~\eqref{eq:dyn} with a dissipative and a diffusive term. 
\begin{comment}
Note that~\eqref{eq:dyn} does not coincide, in general, with the average of Eq.~\eqref{eq:lan}.
Let us also recall that the probability density function (pdf) of the system, $\cP(\vect{x},t)$, evolves through the Fokker-Planck equation
\begin{equation}
\label{eq:fpgeneral}
    \partial_t \cP=\sum_{i}\partial_i\big[(J\vect{\partial}_{\vect{x} } H)_i \cP+\gamma\Gamma_i \cP-\gamma\sum_j \partial_{j} (D_{ij}\cP)\big]\,.
\end{equation}
\end{comment}
The above process describes a system in thermal equilibrium if two conditions are satisfied: (i) the stationary probability density function (pdf) of the problem is the Boltzmann distribution $\cP_S(\vect{x})=Z_{\beta}^{-1}e^{-\beta H(\vect{x})}$, where $\beta$ has the physical meaning of an inverse temperature and $Z_{\beta}$ is the partition function; (ii) detailed balance symmetry~\cite{gardiner2009stochastic} holds, i.e.
 \begin{equation}
\label{eq:detbal}
   \cP_S(\vect{x}_1)W_t\cbr{\vect{x}_2 | \vect{x}_1}=\cP_S(\widehat{\vect{x}}_2)W_t\cbr{ \widehat{\vect{x}}_1 | \widehat{\vect{x}}_2 }
\end{equation}
where $W_t\cbr{\vect{x}_2 | \vect{x}_1}$ is the propagator of the dynamics and $\widehat{\vect{x}}_{1,2}={\op}(\vect{x}_{1,2})$.
By requiring (i) we get~\footnote{The most general form admits an additional term $\gamma h(\vect x)$ on the r.h.s. of Eq.~\eqref{eq:detbal}, with quite restrictive symmetry properties. We show in~\cite{SM} that this term vanishes in most physically-inspired applications.}
\begin{equation}
\label{eq:dissipation_final}
   \Gamma_i(\vect{x})=e^{\beta H(\vect{x})}\sum_j \partial_{x_j} \lx[D_{ij} (\vect{x})e^{-\beta H(\vect{x})}\rx]\,,
\end{equation}
which is a generalized form of the Einstein relation.  As per (ii),
% let us notice that, for deterministic dynamics,  $W_t\cbr{\vect{x}_2 | \vect{x}_1}=\delta(\vect{x}_2-S^t \vect{x}_1)$, where $S^t$ is the semigroup that identifies the time evolution: in this case, it can be shown that the existence of a reversing symmetry $\op$ implies~\eqref{eq:detbal}, see~\cite{SM}. When $\gamma > 0$, an additional constraint must be imposed. As
as shown in~\cite{chetrite2008fluctuation,chetrite2008fluctuation2,gawedzki2013fluctuation,o2024geometric,o2024geometric2} (and also detailed in the Supplemental Material~\cite{SM}),
%, once $\mathcal{P}_S$ is chosen to be the Boltzmann distribution, 
Eq.~\eqref{eq:detbal} is equivalent to 
\begin{equation}
\label{eq:detbalfp}
\mathcal{L}_{\widehat{\vect{x}}_2}^{\it{fw}}\sbr{e^{\beta H(\vect x_2)}W_t\cbr{ \vect{x}_1 | \vect{x}_2} }=e^{\beta H(\vect x_2)}\mathcal{L}_{\vect{x}_2}^{\it{bw}}W_t\cbr{  \vect{x}_1 | \vect{x}_2}   \,,
\end{equation}
where $\mathcal{L}_{\vect{x}_2}^{\it{fw}}$ and $\mathcal{L}_{\widehat{\vect{x}}_2}^{\it{bw}}$ are the forward and backward Fokker-Planck operators, respectively.
Defining $M_{ij}(\vect{x}) \equiv \partial_{x_j} \op_i(\vect x)$, one  finds the condition~\cite{chetrite2008fluctuation,chetrite2008fluctuation2, SM}
\begin{equation}	
\label{eq:mdm}
M(\vect{x})D(\vect{x})M^T(\vect{x})=D(\widehat{\vect{x}})\,.
\end{equation}
Requirements~\eqref{eq:dissipation_final} and~\eqref{eq:mdm} leave much freedom in the choice of $\Gamma$ and $D$, defining a class of equilibrium stochastic processes compatible with the deterministic limit~\eqref{eq:dyn}.

 Consider, as an example, the case of mechanical systems: here the state $\vect{x}=(\vect{q}, \vect{p})\in \mathbb{R}^{2n}$ is a set of canonical coordinates, the conserved energy $H(\vect{q},\vect{p})$ is the Hamiltonian, and 
 \begin{equation}
     f(\vect{x})=J \nabla H(\vect{q},\vect{p})\,,  \quad \text{with}\quad J=\begin{pmatrix}
         0 & I_n \\ -I_n & 0
     \end{pmatrix}\,,
 \end{equation}
 $I_n$ being the identity in $\mathbb{R}^n$.
  If $H(\vect{q},\vect{p})$ exhibits an even dependence on $\vect{p}$ (e.g., in systems with quadratic kinetic terms), then the ${\op}$ operator enforces the usual parity rules ${\op} (\vect{q},\vect{p}) = (\vect{q},-\vect{p})$. The corresponding Klein-Kramers stochastic differential equations are obtained by defining
\begin{equation}
\label{eq:kk}
\Gamma(\vect{x})=\begin{pmatrix}
    0\\
    - \vect{p}
\end{pmatrix}
\quad \quad 
D(\vect{x})=\begin{pmatrix}
    0 &0\\
    0 &1/\beta
\end{pmatrix}\,.   
\end{equation}
The constant $\gamma$ appearing in Eqs.~\eqref{eq:lan} and~\eqref{eq:noisecorr} has the meaning of a damping coefficient. It is easy to verify that Eqs.~\eqref{eq:dissipation_final} and~\eqref{eq:mdm} are fulfilled in this case. This is also true when non-quadratic kinetic terms are considered~\cite{baldovin2018langevin,baldovin2019derivation}. 
A limit case of interest is that of overdamped dynamics, where $f(\vect{x})=0$: for these systems, the identity transformation ${\op}(\vect{x})=\vect{x}$  trivially fulfills~\eqref{eq:reversing_symmetry} and~\eqref{eq:mdm}. Other examples are discussed in~\cite{SM}.

\begin{comment}
One may wonder whether the inversion operator $\op$, which was determined heuristically in the previous examples, can be identified more systematically. The answer turns out to be positive, at least for the class of integrable models, which includes, remarkably, all stable Hamiltonian systems in 1 dimension. We restrict here to the latter case for simplicity, and we also assume that $H$ has a single minimum, but the discussion can be easily generalized.
\end{comment}

\textit{Multi-scale dynamics at equilibrium --- } Consider a slow variable $\vect{x}$, with TR symmetry ${\op}$, at equilibrium with a fast deterministic bath $\vect{y}$, characterized by the TR operator $\op_b$. The first main result of this Letter is that the marginal dynamics of $\vect{x}$ can be expressed in the form~\eqref{eq:lan}
%, reducing to~\eqref{eq:dyn} in the $\gamma \to 0$ limit,
and obeys conditions~\eqref{eq:dissipation_final} and~\eqref{eq:mdm}. More precisely, inspired by~\cite{pavliotis2008multiscale, kelly2017deterministic}, we consider the dynamical system
\begin{equation}
    \begin{cases}
        \dot{\vect{x}} = f(\vect{x})+\lambda\sqrt{\gamma} \,g(\vect{x},\vect{y})\,;\\
        \dot{\vect{y}} = \lambda^2 f_b(\vect{y})+\lambda\sqrt{\gamma} \,g_b(\vect{x},\vect{y})\,.\\
    \end{cases}
    \label{eq:det-model-sigma}
\end{equation}
 Time-scale separation occurs for $\lambda \gg 1$. The parameter $\gamma\ge0$ acts as a coupling constant between $\vect{x}$ and $\vect{y}$. Let us define $H(\vect{x})$ and $H_b(\vect{y})$ as two conserved quantities of the dynamics of $\vect{x}(t)$ and $\vect{y}(t)$ when the two systems are uncoupled ($\gamma=0$). We interpret them as the energies of the subsystems.  When the two systems are coupled, the total energy $H_{\text{tot}}(\vect{x},\vect{y})$ is given by the sum of the energies $H(\vect{x})$ and $ H_b(\vect{y})\gg H(\vect{x})$ that they would have for $\gamma=0$, and by an interaction term $H_{\text{int}}(\vect{x},\vect{y})\ll H_b(\vect{y})$. We require the dynamics of the coupled system ($\gamma>0$) and those of the isolated subsystems ($\gamma=0$)  to fulfill: (i) volume preservation
\begin{equation}
\nabla_x\cdot f=\nabla_y\cdot f_b =0\,,\quad\quad\nabla_x\cdot g=-\nabla_y\cdot g_b\,; \label{eq:div-free-ms}
\end{equation}
(ii) conservation of the energy
\begin{subequations}
\begin{eqnarray}
 f \cdot \nabla_x H = f_b \cdot \nabla_y H_b&=0 \label{eq:en_pres_a}\,,\\
\dot{\vect{x}} \cdot \nabla_x H_{\text{tot}}  +\dot{\vect{y}} \cdot \nabla_y H_{\text{tot}} &=0 \label{eq:en_pres_b}\,;    
\end{eqnarray}
\label{eq:energy-preservation-ms}
\end{subequations}
%where $H_{\text{tot}}(\vect{x},\vect{y})=H(\vect{x})+H_b(\vect{y})+\lambda^{-1} H_{\text{int}}(\vect{x},\vect{y})$;
%where $H_b(\vect{y})$ is the energy of the bath and $H_{\text{tot}}(\vect{x},\vect{y})$ that of the total system;\\
(iii) reversibility under $\op_{\text{tot}}(\vect x,\vect y)=(\op(\vect x),\op_b(\vect y))$.
%Note that conditions~\eqref{eq:div-free-ms}-\eqref{eq:energy-preservation-ms} guarantee that the isolated subsystems are divergence-free and conserve the energies $H_x(\vect{x})$ and $H_y(\vect{y})$.

Under the same ergodicity assumptions required in~\cite{kelly2017deterministic}, one shows that %on the evolution~\eqref{eq:det-model-sigma}
in the limit $\lambda\to\infty$ the dynamics of $\vect{x}$ converges to Eq.~\eqref{eq:lan}, with $\Gamma(\vect{x})$ given by Eq.~\eqref{eq:dissipation_final} and
\begin{align}
&D(\vect x)=\int_0^\infty ds \int d\vect y\,\rho_x(\vect y)\,
g(\vect x,\vect y)g^T(\vect x,\varphi_b^s(\vect y|\vect x))\,\label{eq:pert-diffusion}%\\
%&\Gamma_i(\vect x)=e^{\beta H_x(\vect x)}\sum_j \partial_{x_j} \lx[D_{ij}(\vect x)e^{-\beta H_x(\vect x)}\rx]
\end{align}
where $\rho_x(\vect y)$ and $\varphi_b^s(\vect y|\vect x)$ denote the invariant (microcanonical) distribution and the flow of the fast dynamics at fixed $\vect{x}$. A straightforward calculation shows then that the symmetry~\eqref{eq:mdm} holds for $D$ (independently of $\op_b$). The derivation is detailed in~\cite{SM}.
%Since by construction $f_I(\op_x (\vect x),\op_y (\vect y))=-M_x(\vect x)f_I(\vect x,\vect y)$, $\rho_{\op_x x}(\op_y \vect y)=\rho_{ x}( \vect y) $, $\phi_{\op_x x}^s(\op (\vect y))=\op \cbr{ \phi_{x}^{-s}(\vect y)}$ and the measure $\rho_{ x}( \vect y)d\vect y$ is invariant under the flow, it follows that the symmetry~\eqref{eq:mdm} for $D$ holds (see SM~\cite{} for details). \textbf{FORSE SI PUÒ SOLO DIRE}
\quad\\

\textit{Reversibility in 2D Langevin processes --- } In the light of the above, to establish whether a given Langevin dynamics describes the evolution of a variable at equilibrium with a fast thermal bath, one should first identify the deterministic limit~\eqref{eq:dyn}, then determine the TR operator ${\op}$ fulfilling~\eqref{eq:reversing_symmetry} (if any), and finally
check that conditions~\eqref{eq:dissipation_final} and~\eqref{eq:mdm}  hold. The second step, in particular, does not admit an automatic procedure, and it usually relies on physical intuition. In what follows we provide a criterion, valid for 2D Langevin dynamics, to determine if the process is at equilibrium in the previously specified sense, even if the time reversal operator $\op$ is not \textit{a priori} known. This is our second main result.

Consider a 2D Langevin process 
$$
\dot{\vect{x}} =A(\vect{x})+\vect{\xi}(t)\,,\quad \av{\vect{\xi}(t)\vect{\xi}(t')}=2\gamma D(\vect{x})\delta(t-t')\,.
$$
We observe that, if equilibrium holds, one can introduce an energy related to the stationary pdf by
$$
H(\vect{x})=-\beta^{-1}\ln \cP_S(\vect{x}) -  \beta^{-1}\ln Z_{\beta}\,.
$$
It is therefore uniquely determined by  $\cP_S$, apart from a multiplicative factor and an additive term, which are irrelevant in what follows: indeed, it is immediate to show that the TR symmetry $\op$ of $H$, if it exists, is the same as that of $\widetilde{H}(\vect{x})=- \ln \cP_S(\vect{x})$. $H(\vect{x})$ must be a conserved quantity in the deterministic limit. This may happen in two cases: either $f(\vect{x})=0$ (overdamped limit), or  $f(\vect{x})=J \nabla H(q,p)$ (symplectic structure). We will have to separately test the two possibilities.

We first define $\Gamma(\vect{x})$ according to Eq.~\eqref{eq:dissipation_final}, and compute $f(\vect{x})=A(\vect{x})-\gamma \Gamma(\vect{x})$. If $f(\vect{x})=0$, then ${\op}(\vect{x})=\vect{x}$ fulfills requirement~\eqref{eq:mdm}. The system is therefore at equilibrium and the dynamics is of the overdamped type.
If $f(\vect{x})\ne 0$, the process may still represents a reversible dynamics with a non-trivial deterministic limit  $f(\vect{x})=J \nabla H(q,p)$, where reversible currents are present. To test this possibility, we first check if $f(\vect{x})\cdot\nabla H=0$. If this is the case, we then introduce the canonical transformation $(\phi, I) = \mathcal W^{-1} (q,p)$, leading to action-angle variables. We can perform this transformation because any symplectic dynamics in $\mathbb{R}^2$ is integrable~\cite{arnold}.
%We assume that $H$ has a single minimum, but the discussion can be easily generalized.
$\mathcal W^{-1}$ defines a generalized momentum (action)
\begin{equation}
\label{eq:action}
I(E)=\frac{1}{2 \pi}\oint_{\mathcal{C}(E)} p\, dq\,, 
\end{equation}
where $\mathcal{C}(E)$ is the closed path at constant energy $H(q,p)=E$ in phase space. The new Hamiltonian $K(I)$, obtained by inverting relation~\eqref{eq:action} between energy and action, only depends on $I$. One has therefore
\begin{equation}
\label{eq:detaa}
    \dot{\phi}=\partial_I K(I)\,,\quad    \dot{I}=0\,,
\end{equation}
where the generalized position (angle) $\phi\in[0,2\pi)$ is a cyclic variable. This angle is uniquely determined by choosing a curve $\phi(q,p)=0$ that crosses $\mathcal{C}(E)$ perpendicularly, once for each value of $E$, and then integrating~\eqref{eq:detaa}.  In this new $(\phi, I)$ space, any involution 
\begin{equation}
    \widetilde{\op}(\phi,I)=( \phi_0(I)-\phi ,I)\,,
\end{equation}
where $\phi_0(I)$ is an arbitrary function of $I$,
fulfills~\eqref{eq:reversing_symmetry}, and it trivially preserves the Hamiltonian $K(I)$: it is therefore a candidate TR operator. Its Jacobian matrix reads
\begin{equation}
\widetilde{M}=\begin{pmatrix}
-1 &\partial_I \phi_0\\
0 &1
\end{pmatrix}  \,. 
\end{equation}
Recalling Ito's formula, the diffusion matrix of the dynamics in this space can be written as $\widetilde{D}=WDW^{-1}$ where $W$ is the Jacobian matrix of the $\mathcal{W}$ transformation.
At this point, to determine whether the original process describes an equilibrium underdamped dynamics with $\widetilde{\op}$, we only need to check the validity of Eq.~\eqref{eq:mdm}, namely $\widetilde{M}(\vect{x})\widetilde{D}(\vect{x})\widetilde{M}^T(\vect{x})=\widetilde{D}(\op(\vect{x}))$. This is an overdetermined system of equations for $\phi_0$ and $\partial_I \phi_0$, which in general has no solution. When it does, it leads to a differential equation for $\phi_0$, whose solution is in turn the TR operator under which the process is at equilibrium. Let us notice that the TR operator in the original $(q,p)$ space reads
\begin{equation}
\label{eq:epsaa}
    {\op}(q,p) \equiv \mathcal W (\,\widetilde{\op}(\,\mathcal W^{-1} (q,p)))\,;
\end{equation}
indeed, $\mathcal{W}$ performs a change of basis to action-angle variables, and $\widetilde{\op}$ enforces the TR transformation.

\begin{figure}
    \centering
    \includegraphics[width=0.9\linewidth]{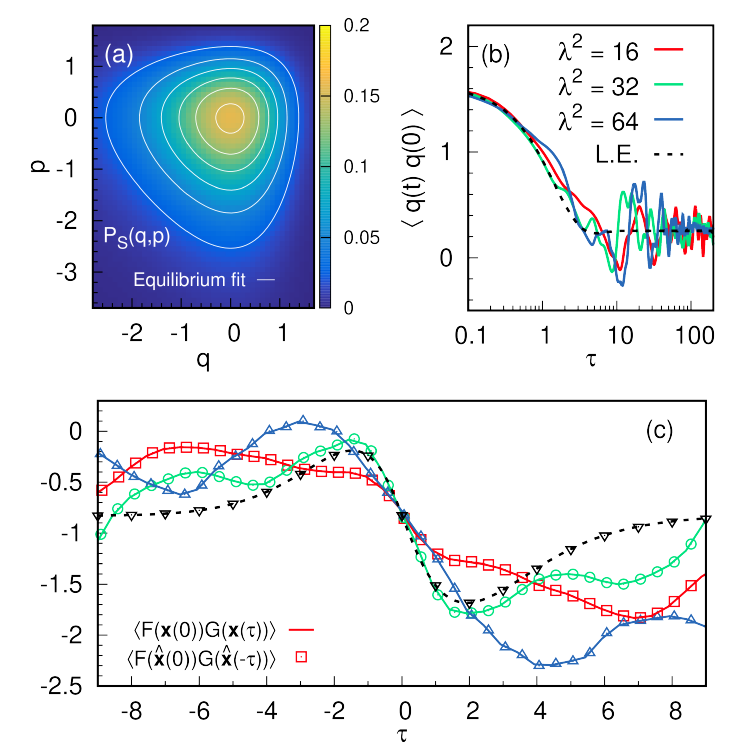}
    \caption{Stochastic representation and TR symmetry of a slow LV process, coupled to many faster dynamics as in Eq.~\eqref{eq:det-model-sigma} (simulation details in the End Matter). Panel (a): stationary distribution (here $\lambda=8$), and its fit with an equilibrium Boltzmann pdf. Panel (b): correlation of the $q(t)$ variable for different values of $\lambda$, and comparison with the expected Langevin Equation in the limit $\lambda\gg1$. Panel (c): check of Eq.~\eqref{eq:dbcorr} in the deterministic evolution for different values of $\lambda$ and for the limit Langevin Equation, with $F(\vect{x})=q$, $G(\vect{x})=p^2$. Color code as in panel (b). }
    \label{fig:lvdet}
\end{figure}

\textit{The prey-predator model --- } 
To test our results, we consider a system of interacting Lotka-Volterra (LV) models. We recall that the prey-predator  LV model~\cite{lotka1920analytical,volterra1926variazioni} reads
\begin{subequations}
\label{eq:lvoriginal}
    \begin{eqnarray}
       \dot{z_1}&=&a z_1 - b z_1z_2\\
       \dot{z_2}&=&-c z_2 + d z_1z_2\,,
    \end{eqnarray}
\end{subequations}
where $z_1$ and $z_2$ represent the populations of two interacting  species, and $a$, $b$, $c$,  $d$ are positive parameters.
The model is equivalent to the Hamiltonian system 
\begin{equation}
    \label{eq:lvham}
    H(\vect{x})=H(q,p)=e^p-p-1 + \frac{a}{c}\cbr{e^q-q-1}\,
\end{equation}
where $q=\log \frac{b z_2}{a}$, $p= \log \frac{d z_1}{c}$
and time is rescaled by $c^{-1}$. Here we will focus on the case $a/c=1$: generalizations are discussed in~\cite{SM}. We consider a slow LV dynamics $\vect{x}(t)$, interacting with a bath $\vect y(t)$ composed  of $N$ fast LV systems, as in Eq.~\eqref{eq:det-model-sigma}, where $f(\vect x)=J\nabla H(\vect x)\,.$
The deterministic dynamics of the bath and its interaction with $\vect x$ are inspired by~\cite{goel1971volterra}: we detail them in the End Matter, where we also verify that Eqs~\eqref{eq:div-free-ms},~\eqref{eq:energy-preservation-ms} hold.
In Fig.~\ref{fig:lvdet}(a) we show that, as expected, $P_{S}(\vect x) \propto e^{-\beta H(\vect x)}$. We then use the fitted value of $\beta$ and Eq.~\eqref{eq:pert-diffusion} to infer an equilibrium Langevin Equation of the form~\eqref{eq:lan}, with $\Gamma$ given by Eq.~\eqref{eq:dissipation_final}. One finds that the diffusion matrix is diagonal, with $D_{qq}=D_{pp}$. Figure~\ref{fig:lvdet}(b) shows the comparison of the correlation function $\av{q(t)q(0)}$ in this stochastic model and in the original deterministic dynamics, for different values of $\lambda$. Up to the time scales of the slow process, the agreement is fairly good, as expected.
We now investigate the TR symmetry. Using our criterion it is easy to verify that the operator that  reverses the evolution is $\op(q,p)=(p,q)$. In Fig.~\ref{fig:lvdet}(c) we check this symmetry verifying the identity
\begin{equation}
\label{eq:dbcorr}
\av{F[\vect{x}(0)]\, G[\vect{x}(\tau)]} = \av{ F[\op (\vect{x}(0))]\, G[\op( \vect{x}(-\tau))]}\,
\end{equation}
which holds at equilibrium for generic observables $F(\vect{x})$, $G(\vect{x})$ as a consequence of detailed balance~\eqref{eq:detbal}. As expected from our first main result, the TR symmetry, which holds in the deterministic limit, is not spoiled by the coupling with our deterministic bath.  Let us notice that in both panels (b) and (c), the deterministic and stochastic correlations show deviations on long time scales, likely due to lack of ergodicity in the global system. These effects, however, do not affect the TR symmetries.

\begin{figure}
    \centering
    \includegraphics[width=0.9\linewidth]{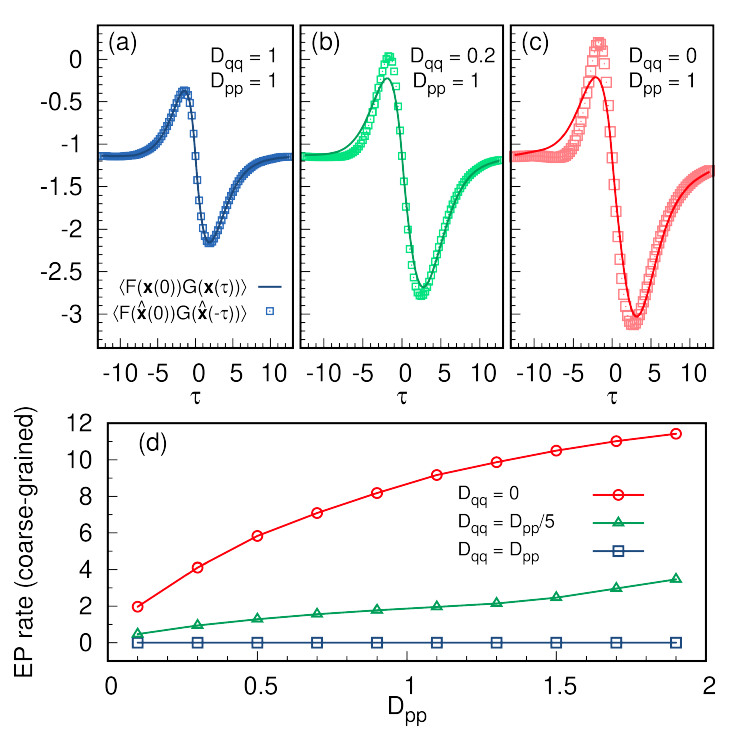}
    \caption{Detailed balance for the stochastic dynamics~\eqref{eq:lan}, where the deterministic limit is the LV model~\eqref{eq:lvham} with $a/c=1$. We test detailed balance through Eq.~\eqref{eq:dbcorr}, with $F(q,p)=p$ and $G(q,p)=q^2$. The l.h.s. (solid line) and r.h.s. (squares) of Eq.~\eqref{eq:dbcorr} are compared. If condition~\eqref{eq:mdm} is fulfilled (panel (a)), detailed balance symmetry holds. If not, as in panels (b) and (c), condition~\eqref{eq:dbcorr} is violated. In panel (d) we show the dependence of the EP rate estimator~\eqref{eq:epest} on the value of $D_p$ in the three cases $D_q=D_p$ (blue squares), $D_q=D_p/5$ (green triangles) and $D_q=0$ (red circles). Only in the first case the entropy production rate vanishes. For the computation of the EP rate, $\sigma=1$, $\tau=0.1$. Here $\beta=1$, $\Delta t =0.005$, $t_{max}=10^6$.}
    \label{fig:lvsymm}
\end{figure}

In the above example we found empirically that, at equilibrium, $D=\text{diag}(D_{qq},D_{pp})$ with $D_{qq}=D_{pp}$. Let us stress that that many other choices of $D$ would lead to a stationary state described by the Boltzmann distribution, but this would not guarantee, in general, that Eq.~\eqref{eq:dbcorr} holds.  To show this point, we perform numerical simulation of a Langevin process~\eqref{eq:lan}, with additive noise $D=\text{diag}(D_{qq},D_{pp})$, and $\Gamma$ given by Eq.~\eqref{eq:dissipation_final}. In Fig.~\ref{fig:lvsymm} we check detailed balance (a) for a choice of $D_{qq}$, $D_{pp}$ that verifies condition~\eqref{eq:mdm} and (b)-(c) for choices that do not. While the pdf (not shown) is the expected equilibrium one in all cases, Fig.~\ref{fig:lvsymm} shows that only condition (a) leads to TR symmetry. Let us stress that condition (c) is formally similar to the structure~\eqref{eq:kk} of Klein-Kramers dynamics, with the thermal noise only affecting $\dot{p}$. Still, in the present case, this choice leads to TR symmetry breaking, since $H(q,p)\ne H(q,-p)$. This point is made even clearer by Fig.~\ref{fig:lvsymm}(d), where we show a coarse-grained proxy of the entropy production (EP) rate, namely
\begin{equation}
\label{eq:epest}
    \dot{\Sigma}_{\tau,\sigma}=\sum_{\vect{x}, \vect{y}\in \Pi(\sigma)}\frac{\mathcal{P}(\vect{x},0;\vect{y},\tau)}{\tau}\ln\cbr{\frac{W(\vect{y},\tau|\vect{x},0)}{W(\op (\vect{x}),\tau| \op (\vect{y}),0)}}\,.
\end{equation}
Here, $\Pi(\sigma)$ is a partition of the $(q,p)$ space through $\sigma \times \sigma$ boxes, while $\tau$ is the time-interval by which we discretize time~\cite{lucente2023statistical,lucente2024conceptual}. The joint and conditioned pdf $\mathcal{P}$ and $W$ are found empirically. $\dot{\Sigma}_{\tau,\sigma}$ is an estimator of detailed-balance breaking at the scales identified by $\sigma$ and $\tau$, and it tends to the EP rate in the limit $\sigma \to 0$, $\tau \to 0$~\cite{lucente2023statistical,lucente2024conceptual}. We observe that $\dot{\Sigma}_{\tau,\sigma}$ vanishes when condition~\eqref{eq:detbal} is fulfilled, while it is consistently larger than zero when it is violated.

\textit{Conclusion --- } In this Letter we have shown that the dynamics of a slow variable coupled to a fast deterministic bath, under physically reasonable assumptions, can be described by a reversible Langevin process that shares the TR symmetry with the original deterministic evolution. This observation leads to a criterion, valid for 2D Langevin equations, to establish whether they can be interpreted as equilibrium dynamics, for a suitable choice of the TR operator. We stress that the actual TR rules can be only determined from the physical intuition on the dynamics: the evolution equations, by themselves, are not enough. However, if the TR operator is not known, our method can be used to exclude that the system is at equilibrium, or, otherwise, to determine under which choice of the TR operator the dynamics is reversible.

In this sense, the analysis presented here is expected to be relevant to the study of out-of-equilibrium systems with a reference equilibrium limit where the usual TR parity rules do not hold, e.g., in the presence of magnetic fields, vorticity, chiral interactions. In future works, the ideas discussed here could be applied to generalize the fluctuation theorems and Onsager's reciprocal relations to the considered class of systems.

\iffalse
In this Letter we have shown that an equilibrium stochastic process at inverse temperature $\beta$ can be obtained by suitably adding diffusion and dissipation to a reversible dynamical system such as, e.g., a stable Hamiltonian model in one dimension, also in the absence of the usual parity rules, $(q,p) \leftrightarrow (q,-p)$. Our physical assumption is that the Fokker-Planck equation describing the evolution of the corresponding pdf should be characterized by the same TR symmetry as its deterministic limit.
Identifying the operator $\op$ that inverts such dynamics  allows us to generalize detailed balance and entropy production to these cases. Without this initial step, these concepts are not properly defined. 

We stress once again that in the case of a stochastic process, $\op$ can be only determined from the physical intuition on the dynamics: the evolution equations, by themselves, are not enough. In this sense, the analysis presented here is expected to be relevant to the study of out-of-equilibrium systems with a reference equilibrium limit where the usual TR parity rules do not hold, e.g., in the presence of magnetic fields, vorticity, chiral interactions. A particularly relevant case is the
LV model, which can be further investigated following the lines of what we presented in this Letter. In future works, the ideas discussed here could be applied to generalize the fluctuation theorems and Onsager's reciprocal relations to the considered class of systems.
\fi

We gratefully thank A. Puglisi and U. Marini Bettolo Marconi for the careful reading of our paper and for their useful suggestions. We also thank S. Melillo, L. Parisi and the COBBS group for hosting our discussions, and for their patience with us.
 The authors acknowledge financial support from the MIUR PRIN 2022 (project ``SNO-MINK'' no. 2022KWTEB7) which is funded by the European Union  Next Generation EU, M4 C2 1.1  CUP B53C24006470006. MB was also supported by ERC Advanced Grant RG.BIO (Contract No. 785932).

\bibliography{biblio}

\begin{thebibliography}{37}%
\makeatletter
\providecommand \@ifxundefined [1]{%
 \@ifx{#1\undefined}
}%
\providecommand \@ifnum [1]{%
 \ifnum #1\expandafter \@firstoftwo
 \else \expandafter \@secondoftwo
 \fi
}%
\providecommand \@ifx [1]{%
 \ifx #1\expandafter \@firstoftwo
 \else \expandafter \@secondoftwo
 \fi
}%
\providecommand \natexlab [1]{#1}%
\providecommand \enquote  [1]{``#1''}%
\providecommand \bibnamefont  [1]{#1}%
\providecommand \bibfnamefont [1]{#1}%
\providecommand \citenamefont [1]{#1}%
\providecommand \href@noop [0]{\@secondoftwo}%
\providecommand \href [0]{\begingroup \@sanitize@url \@href}%
\providecommand \@href[1]{\@@startlink{#1}\@@href}%
\providecommand \@@href[1]{\endgroup#1\@@endlink}%
\providecommand \@sanitize@url [0]{\catcode `\\12\catcode `\$12\catcode
  `\&12\catcode `\#12\catcode `\^12\catcode `\_12\catcode `\%12\relax}%
\providecommand \@@startlink[1]{}%
\providecommand \@@endlink[0]{}%
\providecommand \url  [0]{\begingroup\@sanitize@url \@url }%
\providecommand \@url [1]{\endgroup\@href {#1}{\urlprefix }}%
\providecommand \urlprefix  [0]{URL }%
\providecommand \Eprint [0]{\href }%
\providecommand \doibase [0]{https://doi.org/}%
\providecommand \selectlanguage [0]{\@gobble}%
\providecommand \bibinfo  [0]{\@secondoftwo}%
\providecommand \bibfield  [0]{\@secondoftwo}%
\providecommand \translation [1]{[#1]}%
\providecommand \BibitemOpen [0]{}%
\providecommand \bibitemStop [0]{}%
\providecommand \bibitemNoStop [0]{.\EOS\space}%
\providecommand \EOS [0]{\spacefactor3000\relax}%
\providecommand \BibitemShut  [1]{\csname bibitem#1\endcsname}%
\let\auto@bib@innerbib\@empty
%</preamble>
\bibitem [{\citenamefont {Dieball}\ and\ \citenamefont
  {Godec}(2025)}]{dieball2025perspective}%
  \BibitemOpen
  \bibfield  {author} {\bibinfo {author} {\bibfnamefont {C.}~\bibnamefont
  {Dieball}}\ and\ \bibinfo {author} {\bibfnamefont {A.}~\bibnamefont
  {Godec}},\ }\bibfield  {title} {\bibinfo {title} {Perspective: Time
  irreversibility in systems observed at coarse resolution},\ }\href@noop {}
  {\bibfield  {journal} {\bibinfo  {journal} {The Journal of Chemical Physics}\
  }\textbf {\bibinfo {volume} {162}} (\bibinfo {year} {2025})}\BibitemShut
  {NoStop}%
\bibitem [{\citenamefont {Lebowitz}\ and\ \citenamefont
  {Spohn}(1999)}]{lebowitz1999gallavotti}%
  \BibitemOpen
  \bibfield  {author} {\bibinfo {author} {\bibfnamefont {J.~L.}\ \bibnamefont
  {Lebowitz}}\ and\ \bibinfo {author} {\bibfnamefont {H.}~\bibnamefont
  {Spohn}},\ }\bibfield  {title} {\bibinfo {title} {A
  {Gallavotti}--{Cohen}-type symmetry in the large deviation functional for
  stochastic dynamics},\ }\href@noop {} {\bibfield  {journal} {\bibinfo
  {journal} {Journal of Statistical Physics}\ }\textbf {\bibinfo {volume}
  {95}},\ \bibinfo {pages} {333} (\bibinfo {year} {1999})}\BibitemShut
  {NoStop}%
\bibitem [{\citenamefont {Maes}\ \emph {et~al.}(2000)\citenamefont {Maes},
  \citenamefont {Redig},\ and\ \citenamefont {Moffaert}}]{maes2000definition}%
  \BibitemOpen
  \bibfield  {author} {\bibinfo {author} {\bibfnamefont {C.}~\bibnamefont
  {Maes}}, \bibinfo {author} {\bibfnamefont {F.}~\bibnamefont {Redig}},\ and\
  \bibinfo {author} {\bibfnamefont {A.~V.}\ \bibnamefont {Moffaert}},\
  }\bibfield  {title} {\bibinfo {title} {On the definition of entropy
  production, via examples},\ }\href@noop {} {\bibfield  {journal} {\bibinfo
  {journal} {Journal of Mathematical Physics}\ }\textbf {\bibinfo {volume}
  {41}},\ \bibinfo {pages} {1528} (\bibinfo {year} {2000})}\BibitemShut
  {NoStop}%
\bibitem [{\citenamefont {Gallavotti}\ and\ \citenamefont
  {Cohen}(1995)}]{gallavotti1995dynamical}%
  \BibitemOpen
  \bibfield  {author} {\bibinfo {author} {\bibfnamefont {G.}~\bibnamefont
  {Gallavotti}}\ and\ \bibinfo {author} {\bibfnamefont {E.~G.~D.}\ \bibnamefont
  {Cohen}},\ }\bibfield  {title} {\bibinfo {title} {Dynamical ensembles in
  nonequilibrium statistical mechanics},\ }\href@noop {} {\bibfield  {journal}
  {\bibinfo  {journal} {Physical Review Letters}\ }\textbf {\bibinfo {volume}
  {74}},\ \bibinfo {pages} {2694} (\bibinfo {year} {1995})}\BibitemShut
  {NoStop}%
\bibitem [{\citenamefont {Kurchan}(1998)}]{kurchan1998fluctuation}%
  \BibitemOpen
  \bibfield  {author} {\bibinfo {author} {\bibfnamefont {J.}~\bibnamefont
  {Kurchan}},\ }\bibfield  {title} {\bibinfo {title} {Fluctuation theorem for
  stochastic dynamics},\ }\href@noop {} {\bibfield  {journal} {\bibinfo
  {journal} {Journal of Physics A: Mathematical and General}\ }\textbf
  {\bibinfo {volume} {31}},\ \bibinfo {pages} {3719} (\bibinfo {year}
  {1998})}\BibitemShut {NoStop}%
\bibitem [{\citenamefont {Maes}(2003)}]{maes2003origin}%
  \BibitemOpen
  \bibfield  {author} {\bibinfo {author} {\bibfnamefont {C.}~\bibnamefont
  {Maes}},\ }\bibfield  {title} {\bibinfo {title} {On the origin and the use of
  fluctuation relations for the entropy},\ }\href@noop {} {\bibfield  {journal}
  {\bibinfo  {journal} {S{\'e}minaire Poincar{\'e}}\ }\textbf {\bibinfo
  {volume} {2}},\ \bibinfo {pages} {29} (\bibinfo {year} {2003})}\BibitemShut
  {NoStop}%
\bibitem [{\citenamefont {Speck}\ and\ \citenamefont
  {Seifert}(2005)}]{speck2005integral}%
  \BibitemOpen
  \bibfield  {author} {\bibinfo {author} {\bibfnamefont {T.}~\bibnamefont
  {Speck}}\ and\ \bibinfo {author} {\bibfnamefont {U.}~\bibnamefont
  {Seifert}},\ }\bibfield  {title} {\bibinfo {title} {Integral fluctuation
  theorem for the housekeeping heat},\ }\href@noop {} {\bibfield  {journal}
  {\bibinfo  {journal} {Journal of Physics A: Mathematical and General}\
  }\textbf {\bibinfo {volume} {38}},\ \bibinfo {pages} {L581} (\bibinfo {year}
  {2005})}\BibitemShut {NoStop}%
\bibitem [{\citenamefont {Lamb}\ and\ \citenamefont
  {Roberts}(1998)}]{lamb1998time}%
  \BibitemOpen
  \bibfield  {author} {\bibinfo {author} {\bibfnamefont {J.~S.~W.}\
  \bibnamefont {Lamb}}\ and\ \bibinfo {author} {\bibfnamefont {J.~A.~G.}\
  \bibnamefont {Roberts}},\ }\bibfield  {title} {\bibinfo {title}
  {Time-reversal symmetry in dynamical systems: a survey},\ }\href@noop {}
  {\bibfield  {journal} {\bibinfo  {journal} {Physica D: Nonlinear Phenomena}\
  }\textbf {\bibinfo {volume} {112}},\ \bibinfo {pages} {1} (\bibinfo {year}
  {1998})}\BibitemShut {NoStop}%
\bibitem [{\citenamefont {Lucente}\ \emph {et~al.}(2022)\citenamefont
  {Lucente}, \citenamefont {Baldassarri}, \citenamefont {Puglisi},
  \citenamefont {Vulpiani},\ and\ \citenamefont
  {Viale}}]{lucente2022inference}%
  \BibitemOpen
  \bibfield  {author} {\bibinfo {author} {\bibfnamefont {D.}~\bibnamefont
  {Lucente}}, \bibinfo {author} {\bibfnamefont {A.}~\bibnamefont
  {Baldassarri}}, \bibinfo {author} {\bibfnamefont {A.}~\bibnamefont
  {Puglisi}}, \bibinfo {author} {\bibfnamefont {A.}~\bibnamefont {Vulpiani}},\
  and\ \bibinfo {author} {\bibfnamefont {M.}~\bibnamefont {Viale}},\ }\bibfield
   {title} {\bibinfo {title} {Inference of time irreversibility from incomplete
  information: Linear systems and its pitfalls},\ }\href@noop {} {\bibfield
  {journal} {\bibinfo  {journal} {Physical Review Research}\ }\textbf {\bibinfo
  {volume} {4}},\ \bibinfo {pages} {043103} (\bibinfo {year}
  {2022})}\BibitemShut {NoStop}%
\bibitem [{\citenamefont {Lucente}\ \emph {et~al.}(2025)\citenamefont
  {Lucente}, \citenamefont {Baldovin}, \citenamefont {Cecconi}, \citenamefont
  {Cencini}, \citenamefont {Cocciaglia}, \citenamefont {Puglisi}, \citenamefont
  {Viale},\ and\ \citenamefont {Vulpiani}}]{lucente2024conceptual}%
  \BibitemOpen
  \bibfield  {author} {\bibinfo {author} {\bibfnamefont {D.}~\bibnamefont
  {Lucente}}, \bibinfo {author} {\bibfnamefont {M.}~\bibnamefont {Baldovin}},
  \bibinfo {author} {\bibfnamefont {F.}~\bibnamefont {Cecconi}}, \bibinfo
  {author} {\bibfnamefont {M.}~\bibnamefont {Cencini}}, \bibinfo {author}
  {\bibfnamefont {N.}~\bibnamefont {Cocciaglia}}, \bibinfo {author}
  {\bibfnamefont {A.}~\bibnamefont {Puglisi}}, \bibinfo {author} {\bibfnamefont
  {M.}~\bibnamefont {Viale}},\ and\ \bibinfo {author} {\bibfnamefont
  {A.}~\bibnamefont {Vulpiani}},\ }\bibfield  {title} {\bibinfo {title}
  {Conceptual and practical approaches for investigating irreversible
  processes},\ }\href {https://doi.org/10.1088/1367-2630/adc6ab} {\bibfield
  {journal} {\bibinfo  {journal} {New Journal of Physics}\ }\textbf {\bibinfo
  {volume} {27}},\ \bibinfo {pages} {041201} (\bibinfo {year}
  {2025})}\BibitemShut {NoStop}%
\bibitem [{\citenamefont {Loos}\ and\ \citenamefont
  {Klapp}(2020)}]{loos2020irreversibility}%
  \BibitemOpen
  \bibfield  {author} {\bibinfo {author} {\bibfnamefont {S.~A.}\ \bibnamefont
  {Loos}}\ and\ \bibinfo {author} {\bibfnamefont {S.~H.}\ \bibnamefont
  {Klapp}},\ }\bibfield  {title} {\bibinfo {title} {Irreversibility, heat and
  information flows induced by non-reciprocal interactions},\ }\href@noop {}
  {\bibfield  {journal} {\bibinfo  {journal} {New Journal of Physics}\ }\textbf
  {\bibinfo {volume} {22}},\ \bibinfo {pages} {123051} (\bibinfo {year}
  {2020})}\BibitemShut {NoStop}%
\bibitem [{\citenamefont {Krapivsky}\ \emph {et~al.}(2010)\citenamefont
  {Krapivsky}, \citenamefont {Redner},\ and\ \citenamefont
  {Ben-Naim}}]{krapivsky}%
  \BibitemOpen
  \bibfield  {author} {\bibinfo {author} {\bibfnamefont {P.~L.}\ \bibnamefont
  {Krapivsky}}, \bibinfo {author} {\bibfnamefont {S.}~\bibnamefont {Redner}},\
  and\ \bibinfo {author} {\bibfnamefont {E.}~\bibnamefont {Ben-Naim}},\
  }\href@noop {} {\emph {\bibinfo {title} {A Kinetic View of Statistical
  Physics}}}\ (\bibinfo  {publisher} {Cambridge University Press},\ \bibinfo
  {address} {Cambridge, UK},\ \bibinfo {year} {2010})\BibitemShut {NoStop}%
\bibitem [{\citenamefont {van Kampen}(1997)}]{10.1007/BFb0105593}%
  \BibitemOpen
  \bibfield  {author} {\bibinfo {author} {\bibfnamefont {N.~G.}\ \bibnamefont
  {van Kampen}},\ }\bibfield  {title} {\bibinfo {title} {Probability in
  physics},\ }in\ \href@noop {} {\emph {\bibinfo {booktitle} {Stochastic
  Dynamics}}},\ \bibinfo {editor} {edited by\ \bibinfo {editor} {\bibfnamefont
  {L.}~\bibnamefont {Schimansky-Geier}}\ and\ \bibinfo {editor} {\bibfnamefont
  {T.}~\bibnamefont {P{\"o}schel}}}\ (\bibinfo  {publisher} {Springer Berlin
  Heidelberg},\ \bibinfo {address} {Berlin, Heidelberg},\ \bibinfo {year}
  {1997})\ pp.\ \bibinfo {pages} {1--4}\BibitemShut {NoStop}%
\bibitem [{\citenamefont {Zwanzig}(2001)}]{zwanzig2001nonequilibrium}%
  \BibitemOpen
  \bibfield  {author} {\bibinfo {author} {\bibfnamefont {R.}~\bibnamefont
  {Zwanzig}},\ }\href@noop {} {\emph {\bibinfo {title} {Nonequilibrium
  statistical mechanics}}}\ (\bibinfo  {publisher} {Oxford University Press},\
  \bibinfo {year} {2001})\BibitemShut {NoStop}%
\bibitem [{\citenamefont {Cépas}\ and\ \citenamefont
  {Kurchan}(1998)}]{Cpas1998}%
  \BibitemOpen
  \bibfield  {author} {\bibinfo {author} {\bibfnamefont {O.}~\bibnamefont
  {Cépas}}\ and\ \bibinfo {author} {\bibfnamefont {J.}~\bibnamefont
  {Kurchan}},\ }\bibfield  {title} {\bibinfo {title} {Canonically invariant
  formulation of {Langevin} and {Fokker}-{Planck} equations},\ }\href
  {https://doi.org/10.1007/s100510050243} {\bibfield  {journal} {\bibinfo
  {journal} {The European Physical Journal B}\ }\textbf {\bibinfo {volume}
  {2}},\ \bibinfo {pages} {221–223} (\bibinfo {year} {1998})}\BibitemShut
  {NoStop}%
\bibitem [{\citenamefont {Guioth}\ \emph {et~al.}(2022)\citenamefont {Guioth},
  \citenamefont {Bouchet},\ and\ \citenamefont {Eyink}}]{guioth2022path}%
  \BibitemOpen
  \bibfield  {author} {\bibinfo {author} {\bibfnamefont {J.}~\bibnamefont
  {Guioth}}, \bibinfo {author} {\bibfnamefont {F.}~\bibnamefont {Bouchet}},\
  and\ \bibinfo {author} {\bibfnamefont {G.~L.}\ \bibnamefont {Eyink}},\
  }\bibfield  {title} {\bibinfo {title} {Path large deviations for the kinetic
  theory of weak turbulence},\ }\href@noop {} {\bibfield  {journal} {\bibinfo
  {journal} {Journal of Statistical Physics}\ }\textbf {\bibinfo {volume}
  {189}},\ \bibinfo {pages} {20} (\bibinfo {year} {2022})}\BibitemShut
  {NoStop}%
\bibitem [{\citenamefont {Bertini}\ \emph {et~al.}(2015)\citenamefont
  {Bertini}, \citenamefont {De~Sole}, \citenamefont {Gabrielli}, \citenamefont
  {Jona-Lasinio},\ and\ \citenamefont {Landim}}]{bertini2015macroscopic}%
  \BibitemOpen
  \bibfield  {author} {\bibinfo {author} {\bibfnamefont {L.}~\bibnamefont
  {Bertini}}, \bibinfo {author} {\bibfnamefont {A.}~\bibnamefont {De~Sole}},
  \bibinfo {author} {\bibfnamefont {D.}~\bibnamefont {Gabrielli}}, \bibinfo
  {author} {\bibfnamefont {G.}~\bibnamefont {Jona-Lasinio}},\ and\ \bibinfo
  {author} {\bibfnamefont {C.}~\bibnamefont {Landim}},\ }\bibfield  {title}
  {\bibinfo {title} {Macroscopic fluctuation theory},\ }\href@noop {}
  {\bibfield  {journal} {\bibinfo  {journal} {Reviews of Modern Physics}\
  }\textbf {\bibinfo {volume} {87}},\ \bibinfo {pages} {593} (\bibinfo {year}
  {2015})}\BibitemShut {NoStop}%
\bibitem [{\citenamefont {Falasco}\ and\ \citenamefont
  {Esposito}(2025)}]{falasco2025macroscopic}%
  \BibitemOpen
  \bibfield  {author} {\bibinfo {author} {\bibfnamefont {G.}~\bibnamefont
  {Falasco}}\ and\ \bibinfo {author} {\bibfnamefont {M.}~\bibnamefont
  {Esposito}},\ }\bibfield  {title} {\bibinfo {title} {Macroscopic stochastic
  thermodynamics},\ }\href@noop {} {\bibfield  {journal} {\bibinfo  {journal}
  {Reviews of Modern Physics}\ }\textbf {\bibinfo {volume} {97}},\ \bibinfo
  {pages} {015002} (\bibinfo {year} {2025})}\BibitemShut {NoStop}%
\bibitem [{\citenamefont {Pavliotis}\ and\ \citenamefont
  {Stuart}(2008)}]{pavliotis2008multiscale}%
  \BibitemOpen
  \bibfield  {author} {\bibinfo {author} {\bibfnamefont {G.}~\bibnamefont
  {Pavliotis}}\ and\ \bibinfo {author} {\bibfnamefont {A.}~\bibnamefont
  {Stuart}},\ }\href@noop {} {\emph {\bibinfo {title} {Multiscale methods:
  averaging and homogenization}}}\ (\bibinfo  {publisher} {Springer Science \&
  Business Media},\ \bibinfo {year} {2008})\BibitemShut {NoStop}%
\bibitem [{\citenamefont {Bonella}\ \emph {et~al.}(2015)\citenamefont
  {Bonella}, \citenamefont {Ciccotti},\ and\ \citenamefont
  {Rondoni}}]{bonella2015time}%
  \BibitemOpen
  \bibfield  {author} {\bibinfo {author} {\bibfnamefont {S.}~\bibnamefont
  {Bonella}}, \bibinfo {author} {\bibfnamefont {G.}~\bibnamefont {Ciccotti}},\
  and\ \bibinfo {author} {\bibfnamefont {L.}~\bibnamefont {Rondoni}},\
  }\bibfield  {title} {\bibinfo {title} {Time reversal symmetry in
  time-dependent correlation functions for systems in a constant magnetic
  field},\ }\href@noop {} {\bibfield  {journal} {\bibinfo  {journal}
  {Europhysics Letters}\ }\textbf {\bibinfo {volume} {108}},\ \bibinfo {pages}
  {60004} (\bibinfo {year} {2015})}\BibitemShut {NoStop}%
\bibitem [{\citenamefont {Bonella}\ \emph {et~al.}(2017)\citenamefont
  {Bonella}, \citenamefont {Coretti}, \citenamefont {Rondoni},\ and\
  \citenamefont {Ciccotti}}]{bonella2017time}%
  \BibitemOpen
  \bibfield  {author} {\bibinfo {author} {\bibfnamefont {S.}~\bibnamefont
  {Bonella}}, \bibinfo {author} {\bibfnamefont {A.}~\bibnamefont {Coretti}},
  \bibinfo {author} {\bibfnamefont {L.}~\bibnamefont {Rondoni}},\ and\ \bibinfo
  {author} {\bibfnamefont {G.}~\bibnamefont {Ciccotti}},\ }\bibfield  {title}
  {\bibinfo {title} {Time-reversal symmetry for systems in a constant external
  magnetic field},\ }\href@noop {} {\bibfield  {journal} {\bibinfo  {journal}
  {Physical Review E}\ }\textbf {\bibinfo {volume} {96}},\ \bibinfo {pages}
  {012160} (\bibinfo {year} {2017})}\BibitemShut {NoStop}%
\bibitem [{\citenamefont {Chetrite}\ and\ \citenamefont
  {Gawedzki}(2008)}]{chetrite2008fluctuation}%
  \BibitemOpen
  \bibfield  {author} {\bibinfo {author} {\bibfnamefont {R.}~\bibnamefont
  {Chetrite}}\ and\ \bibinfo {author} {\bibfnamefont {K.}~\bibnamefont
  {Gawedzki}},\ }\bibfield  {title} {\bibinfo {title} {Fluctuation relations
  for diffusion processes},\ }\href@noop {} {\bibfield  {journal} {\bibinfo
  {journal} {Communications in Mathematical Physics}\ }\textbf {\bibinfo
  {volume} {282}},\ \bibinfo {pages} {469} (\bibinfo {year}
  {2008})}\BibitemShut {NoStop}%
\bibitem [{\citenamefont {Chetrite}\ \emph {et~al.}(2008)\citenamefont
  {Chetrite}, \citenamefont {Falkovich},\ and\ \citenamefont
  {Gawedzki}}]{chetrite2008fluctuation2}%
  \BibitemOpen
  \bibfield  {author} {\bibinfo {author} {\bibfnamefont {R.}~\bibnamefont
  {Chetrite}}, \bibinfo {author} {\bibfnamefont {G.}~\bibnamefont
  {Falkovich}},\ and\ \bibinfo {author} {\bibfnamefont {K.}~\bibnamefont
  {Gawedzki}},\ }\bibfield  {title} {\bibinfo {title} {Fluctuation relations in
  simple examples of non-equilibrium steady states},\ }\href@noop {} {\bibfield
   {journal} {\bibinfo  {journal} {Journal of Statistical Mechanics: Theory and
  Experiment}\ }\textbf {\bibinfo {volume} {2008}},\ \bibinfo {pages} {P08005}
  (\bibinfo {year} {2008})}\BibitemShut {NoStop}%
\bibitem [{\citenamefont {Gawedzki}(2013)}]{gawedzki2013fluctuation}%
  \BibitemOpen
  \bibfield  {author} {\bibinfo {author} {\bibfnamefont {K.}~\bibnamefont
  {Gawedzki}},\ }\bibfield  {title} {\bibinfo {title} {Fluctuation relations in
  stochastic thermodynamics},\ }\href@noop {} {\bibfield  {journal} {\bibinfo
  {journal} {arXiv preprint arXiv:1308.1518}\ } (\bibinfo {year}
  {2013})}\BibitemShut {NoStop}%
\bibitem [{\citenamefont {O’Byrne}\ and\ \citenamefont
  {Cates}(2024)}]{o2024geometric}%
  \BibitemOpen
  \bibfield  {author} {\bibinfo {author} {\bibfnamefont {J.}~\bibnamefont
  {O’Byrne}}\ and\ \bibinfo {author} {\bibfnamefont {M.}~\bibnamefont
  {Cates}},\ }\bibfield  {title} {\bibinfo {title} {Geometric theory of
  (extended) time-reversal symmetries in stochastic processes: I. finite
  dimension},\ }\href@noop {} {\bibfield  {journal} {\bibinfo  {journal}
  {Journal of Statistical Mechanics: Theory and Experiment}\ }\textbf {\bibinfo
  {volume} {2024}},\ \bibinfo {pages} {113207} (\bibinfo {year}
  {2024})}\BibitemShut {NoStop}%
\bibitem [{\citenamefont {O'Byrne}\ and\ \citenamefont
  {Cates}(2024)}]{o2024geometric2}%
  \BibitemOpen
  \bibfield  {author} {\bibinfo {author} {\bibfnamefont {J.}~\bibnamefont
  {O'Byrne}}\ and\ \bibinfo {author} {\bibfnamefont {M.~E.}\ \bibnamefont
  {Cates}},\ }\bibfield  {title} {\bibinfo {title} {Geometric theory of
  (extended) time-reversal symmetries in stochastic processes--part ii: field
  theory},\ }\href@noop {} {\bibfield  {journal} {\bibinfo  {journal} {arXiv
  preprint arXiv:2411.19299}\ } (\bibinfo {year} {2024})}\BibitemShut {NoStop}%
\bibitem [{\citenamefont {Gardiner}(2009)}]{gardiner2009stochastic}%
  \BibitemOpen
  \bibfield  {author} {\bibinfo {author} {\bibfnamefont {C.}~\bibnamefont
  {Gardiner}},\ }\href@noop {} {\emph {\bibinfo {title} {Stochastic
  methods}}},\ Vol.~\bibinfo {volume} {4}\ (\bibinfo  {publisher} {Springer
  Berlin Heidelberg},\ \bibinfo {year} {2009})\BibitemShut {NoStop}%
\bibitem [{Note1()}]{Note1}%
  \BibitemOpen
  \bibinfo {note} {The most general form admits an additional term $\gamma
  h(\protect \vect x)$ on the r.h.s. of Eq.~\protect \eqref {eq:detbal}, with
  quite restrictive symmetry properties. We show in~\cite {SM} that this term
  vanishes in most physically-inspired applications.}\BibitemShut {Stop}%
\bibitem [{SM()}]{SM}%
  \BibitemOpen
  \href@noop {} {}\bibinfo {note} {Supplemental material for this paper can be
  found at [URL]. It includes: a discussion about reversibility in
  deterministic systems, the derivation of Eqs.~(7),~(9) and~(15), and a series
  of examples with non-trivial time-reversal parity rules.}\BibitemShut {Stop}%
\bibitem [{\citenamefont {Baldovin}\ \emph {et~al.}(2018)\citenamefont
  {Baldovin}, \citenamefont {Puglisi},\ and\ \citenamefont
  {Vulpiani}}]{baldovin2018langevin}%
  \BibitemOpen
  \bibfield  {author} {\bibinfo {author} {\bibfnamefont {M.}~\bibnamefont
  {Baldovin}}, \bibinfo {author} {\bibfnamefont {A.}~\bibnamefont {Puglisi}},\
  and\ \bibinfo {author} {\bibfnamefont {A.}~\bibnamefont {Vulpiani}},\
  }\bibfield  {title} {\bibinfo {title} {{Langevin} equation in systems with
  also negative temperatures},\ }\href@noop {} {\bibfield  {journal} {\bibinfo
  {journal} {Journal of Statistical Mechanics: Theory and Experiment}\ }\textbf
  {\bibinfo {volume} {2018}},\ \bibinfo {pages} {043207} (\bibinfo {year}
  {2018})}\BibitemShut {NoStop}%
\bibitem [{\citenamefont {Baldovin}\ \emph {et~al.}(2019)\citenamefont
  {Baldovin}, \citenamefont {Vulpiani}, \citenamefont {Puglisi},\ and\
  \citenamefont {Prados}}]{baldovin2019derivation}%
  \BibitemOpen
  \bibfield  {author} {\bibinfo {author} {\bibfnamefont {M.}~\bibnamefont
  {Baldovin}}, \bibinfo {author} {\bibfnamefont {A.}~\bibnamefont {Vulpiani}},
  \bibinfo {author} {\bibfnamefont {A.}~\bibnamefont {Puglisi}},\ and\ \bibinfo
  {author} {\bibfnamefont {A.}~\bibnamefont {Prados}},\ }\bibfield  {title}
  {\bibinfo {title} {Derivation of a {Langevin} equation in a system with
  multiple scales: the case of negative temperatures},\ }\href@noop {}
  {\bibfield  {journal} {\bibinfo  {journal} {Physical Review E}\ }\textbf
  {\bibinfo {volume} {99}},\ \bibinfo {pages} {060101} (\bibinfo {year}
  {2019})}\BibitemShut {NoStop}%
\bibitem [{\citenamefont {Kelly}\ and\ \citenamefont
  {Melbourne}(2017)}]{kelly2017deterministic}%
  \BibitemOpen
  \bibfield  {author} {\bibinfo {author} {\bibfnamefont {D.}~\bibnamefont
  {Kelly}}\ and\ \bibinfo {author} {\bibfnamefont {I.}~\bibnamefont
  {Melbourne}},\ }\bibfield  {title} {\bibinfo {title} {Deterministic
  homogenization for fast--slow systems with chaotic noise},\ }\href@noop {}
  {\bibfield  {journal} {\bibinfo  {journal} {Journal of Functional Analysis}\
  }\textbf {\bibinfo {volume} {272}},\ \bibinfo {pages} {4063} (\bibinfo {year}
  {2017})}\BibitemShut {NoStop}%
\bibitem [{\citenamefont {Arnold}(1989)}]{arnold}%
  \BibitemOpen
  \bibfield  {author} {\bibinfo {author} {\bibfnamefont {V.~I.}\ \bibnamefont
  {Arnold}},\ }\href@noop {} {\emph {\bibinfo {title} {Mathematical Methods of
  Classical Mechanics}}}\ (\bibinfo  {publisher} {Springer New York, NY},\
  \bibinfo {year} {1989})\BibitemShut {NoStop}%
\bibitem [{\citenamefont {Lotka}(1920)}]{lotka1920analytical}%
  \BibitemOpen
  \bibfield  {author} {\bibinfo {author} {\bibfnamefont {A.~J.}\ \bibnamefont
  {Lotka}},\ }\bibfield  {title} {\bibinfo {title} {Analytical note on certain
  rhythmic relations in organic systems},\ }\href@noop {} {\bibfield  {journal}
  {\bibinfo  {journal} {Proceedings of the National Academy of Sciences}\
  }\textbf {\bibinfo {volume} {6}},\ \bibinfo {pages} {410} (\bibinfo {year}
  {1920})}\BibitemShut {NoStop}%
\bibitem [{\citenamefont {Volterra}(1926)}]{volterra1926variazioni}%
  \BibitemOpen
  \bibfield  {author} {\bibinfo {author} {\bibfnamefont {V.}~\bibnamefont
  {Volterra}},\ }\href@noop {} {\emph {\bibinfo {title} {Variazioni e
  fluttuazioni del numero d'individui in specie animali conviventi}}}\
  (\bibinfo  {publisher} {Societ{\`a} anonima tipografica ``Leonardo da
  Vinci''},\ \bibinfo {year} {1926})\BibitemShut {NoStop}%
\bibitem [{\citenamefont {Goel}\ \emph {et~al.}(1971)\citenamefont {Goel},
  \citenamefont {Maitra},\ and\ \citenamefont {Montroll}}]{goel1971volterra}%
  \BibitemOpen
  \bibfield  {author} {\bibinfo {author} {\bibfnamefont {N.~S.}\ \bibnamefont
  {Goel}}, \bibinfo {author} {\bibfnamefont {S.~C.}\ \bibnamefont {Maitra}},\
  and\ \bibinfo {author} {\bibfnamefont {E.~W.}\ \bibnamefont {Montroll}},\
  }\bibfield  {title} {\bibinfo {title} {On the {Volterra} and other nonlinear
  models of interacting populations},\ }\href@noop {} {\bibfield  {journal}
  {\bibinfo  {journal} {Reviews of modern physics}\ }\textbf {\bibinfo {volume}
  {43}},\ \bibinfo {pages} {231} (\bibinfo {year} {1971})}\BibitemShut
  {NoStop}%
\bibitem [{\citenamefont {Lucente}\ \emph {et~al.}(2023)\citenamefont
  {Lucente}, \citenamefont {Puglisi}, \citenamefont {Viale},\ and\
  \citenamefont {Vulpiani}}]{lucente2023statistical}%
  \BibitemOpen
  \bibfield  {author} {\bibinfo {author} {\bibfnamefont {D.}~\bibnamefont
  {Lucente}}, \bibinfo {author} {\bibfnamefont {A.}~\bibnamefont {Puglisi}},
  \bibinfo {author} {\bibfnamefont {M.}~\bibnamefont {Viale}},\ and\ \bibinfo
  {author} {\bibfnamefont {A.}~\bibnamefont {Vulpiani}},\ }\bibfield  {title}
  {\bibinfo {title} {Statistical features of systems driven by non-{Gaussian}
  processes: theory \& practice},\ }\href@noop {} {\bibfield  {journal}
  {\bibinfo  {journal} {Journal of Statistical Mechanics: Theory and
  Experiment}\ }\textbf {\bibinfo {volume} {2023}},\ \bibinfo {pages} {113202}
  (\bibinfo {year} {2023})}\BibitemShut {NoStop}%
\end{thebibliography}%


\begin{thebibliography}{14}%
\makeatletter
\providecommand \@ifxundefined [1]{%
 \@ifx{#1\undefined}
}%
\providecommand \@ifnum [1]{%
 \ifnum #1\expandafter \@firstoftwo
 \else \expandafter \@secondoftwo
 \fi
}%
\providecommand \@ifx [1]{%
 \ifx #1\expandafter \@firstoftwo
 \else \expandafter \@secondoftwo
 \fi
}%
\providecommand \natexlab [1]{#1}%
\providecommand \enquote  [1]{``#1''}%
\providecommand \bibnamefont  [1]{#1}%
\providecommand \bibfnamefont [1]{#1}%
\providecommand \citenamefont [1]{#1}%
\providecommand \href@noop [0]{\@secondoftwo}%
\providecommand \href [0]{\begingroup \@sanitize@url \@href}%
\providecommand \@href[1]{\@@startlink{#1}\@@href}%
\providecommand \@@href[1]{\endgroup#1\@@endlink}%
\providecommand \@sanitize@url [0]{\catcode `\\12\catcode `\$12\catcode
  `\&12\catcode `\#12\catcode `\^12\catcode `\_12\catcode `\%12\relax}%
\providecommand \@@startlink[1]{}%
\providecommand \@@endlink[0]{}%
\providecommand \url  [0]{\begingroup\@sanitize@url \@url }%
\providecommand \@url [1]{\endgroup\@href {#1}{\urlprefix }}%
\providecommand \urlprefix  [0]{URL }%
\providecommand \Eprint [0]{\href }%
\providecommand \doibase [0]{https://doi.org/}%
\providecommand \selectlanguage [0]{\@gobble}%
\providecommand \bibinfo  [0]{\@secondoftwo}%
\providecommand \bibfield  [0]{\@secondoftwo}%
\providecommand \translation [1]{[#1]}%
\providecommand \BibitemOpen [0]{}%
\providecommand \bibitemStop [0]{}%
\providecommand \bibitemNoStop [0]{.\EOS\space}%
\providecommand \EOS [0]{\spacefactor3000\relax}%
\providecommand \BibitemShut  [1]{\csname bibitem#1\endcsname}%
\let\auto@bib@innerbib\@empty
%</preamble>
\bibitem [{\citenamefont {Lamb}\ and\ \citenamefont
  {Roberts}(1998)}]{lamb1998time}%
  \BibitemOpen
  \bibfield  {author} {\bibinfo {author} {\bibfnamefont {J.~S.~W.}\
  \bibnamefont {Lamb}}\ and\ \bibinfo {author} {\bibfnamefont {J.~A.~G.}\
  \bibnamefont {Roberts}},\ }\bibfield  {title} {\bibinfo {title}
  {Time-reversal symmetry in dynamical systems: a survey},\ }\href@noop {}
  {\bibfield  {journal} {\bibinfo  {journal} {Physica D: Nonlinear Phenomena}\
  }\textbf {\bibinfo {volume} {112}},\ \bibinfo {pages} {1} (\bibinfo {year}
  {1998})}\BibitemShut {NoStop}%
\bibitem [{\citenamefont {Pavliotis}\ and\ \citenamefont
  {Stuart}(2008)}]{pavliotis2008multiscale}%
  \BibitemOpen
  \bibfield  {author} {\bibinfo {author} {\bibfnamefont {G.}~\bibnamefont
  {Pavliotis}}\ and\ \bibinfo {author} {\bibfnamefont {A.}~\bibnamefont
  {Stuart}},\ }\href@noop {} {\emph {\bibinfo {title} {Multiscale methods:
  averaging and homogenization}}}\ (\bibinfo  {publisher} {Springer Science \&
  Business Media},\ \bibinfo {year} {2008})\BibitemShut {NoStop}%
\bibitem [{\citenamefont {Kelly}\ and\ \citenamefont
  {Melbourne}(2017)}]{kelly2017deterministic}%
  \BibitemOpen
  \bibfield  {author} {\bibinfo {author} {\bibfnamefont {D.}~\bibnamefont
  {Kelly}}\ and\ \bibinfo {author} {\bibfnamefont {I.}~\bibnamefont
  {Melbourne}},\ }\bibfield  {title} {\bibinfo {title} {Deterministic
  homogenization for fast--slow systems with chaotic noise},\ }\href@noop {}
  {\bibfield  {journal} {\bibinfo  {journal} {Journal of Functional Analysis}\
  }\textbf {\bibinfo {volume} {272}},\ \bibinfo {pages} {4063} (\bibinfo {year}
  {2017})}\BibitemShut {NoStop}%
\bibitem [{\citenamefont {Chavanis}(2001)}]{chavanis2001kinetic}%
  \BibitemOpen
  \bibfield  {author} {\bibinfo {author} {\bibfnamefont {P.-H.}\ \bibnamefont
  {Chavanis}},\ }\bibfield  {title} {\bibinfo {title} {Kinetic theory of point
  vortices: diffusion coefficient and systematic drift},\ }\href@noop {}
  {\bibfield  {journal} {\bibinfo  {journal} {Physical Review E}\ }\textbf
  {\bibinfo {volume} {64}},\ \bibinfo {pages} {026309} (\bibinfo {year}
  {2001})}\BibitemShut {NoStop}%
\bibitem [{\citenamefont {Bouchet}\ \emph
  {et~al.}(2014{\natexlab{a}})\citenamefont {Bouchet}, \citenamefont {Laurie},\
  and\ \citenamefont {Zaboronski}}]{bouchet2014langevin}%
  \BibitemOpen
  \bibfield  {author} {\bibinfo {author} {\bibfnamefont {F.}~\bibnamefont
  {Bouchet}}, \bibinfo {author} {\bibfnamefont {J.}~\bibnamefont {Laurie}},\
  and\ \bibinfo {author} {\bibfnamefont {O.}~\bibnamefont {Zaboronski}},\
  }\bibfield  {title} {\bibinfo {title} {{Langevin} dynamics, large deviations
  and instantons for the quasi-geostrophic model and two-dimensional euler
  equations},\ }\href@noop {} {\bibfield  {journal} {\bibinfo  {journal}
  {Journal of Statistical Physics}\ }\textbf {\bibinfo {volume} {156}},\
  \bibinfo {pages} {1066} (\bibinfo {year} {2014}{\natexlab{a}})}\BibitemShut
  {NoStop}%
\bibitem [{\citenamefont {Bouchet}\ \emph
  {et~al.}(2014{\natexlab{b}})\citenamefont {Bouchet}, \citenamefont
  {Nardini},\ and\ \citenamefont {Tangarife}}]{bouchet2014non}%
  \BibitemOpen
  \bibfield  {author} {\bibinfo {author} {\bibfnamefont {F.}~\bibnamefont
  {Bouchet}}, \bibinfo {author} {\bibfnamefont {C.}~\bibnamefont {Nardini}},\
  and\ \bibinfo {author} {\bibfnamefont {T.}~\bibnamefont {Tangarife}},\
  }\bibfield  {title} {\bibinfo {title} {Non-equilibrium statistical mechanics
  of the stochastic {Navier}--{Stokes} equations and geostrophic turbulence},\
  }\href@noop {} {\bibfield  {journal} {\bibinfo  {journal} {5th Warsaw School
  of Statistical Physics}\ } (\bibinfo {year}
  {2014}{\natexlab{b}})}\BibitemShut {NoStop}%
\bibitem [{\citenamefont {Kraichnan}\ and\ \citenamefont
  {Montgomery}(1980)}]{kraichnan1980two}%
  \BibitemOpen
  \bibfield  {author} {\bibinfo {author} {\bibfnamefont {R.~H.}\ \bibnamefont
  {Kraichnan}}\ and\ \bibinfo {author} {\bibfnamefont {D.}~\bibnamefont
  {Montgomery}},\ }\bibfield  {title} {\bibinfo {title} {Two-dimensional
  turbulence},\ }\href@noop {} {\bibfield  {journal} {\bibinfo  {journal}
  {Reports on Progress in Physics}\ }\textbf {\bibinfo {volume} {43}},\
  \bibinfo {pages} {547} (\bibinfo {year} {1980})}\BibitemShut {NoStop}%
\bibitem [{\citenamefont {Majda}\ \emph {et~al.}(2001)\citenamefont {Majda},
  \citenamefont {Timofeyev},\ and\ \citenamefont
  {Vanden~Eijnden}}]{majda2001mathematical}%
  \BibitemOpen
  \bibfield  {author} {\bibinfo {author} {\bibfnamefont {A.~J.}\ \bibnamefont
  {Majda}}, \bibinfo {author} {\bibfnamefont {I.}~\bibnamefont {Timofeyev}},\
  and\ \bibinfo {author} {\bibfnamefont {E.}~\bibnamefont {Vanden~Eijnden}},\
  }\bibfield  {title} {\bibinfo {title} {A mathematical framework for
  stochastic climate models},\ }\href@noop {} {\bibfield  {journal} {\bibinfo
  {journal} {Communications on Pure and Applied Mathematics: A Journal Issued
  by the Courant Institute of Mathematical Sciences}\ }\textbf {\bibinfo
  {volume} {54}},\ \bibinfo {pages} {891} (\bibinfo {year} {2001})}\BibitemShut
  {NoStop}%
\bibitem [{\citenamefont {Boffetta}\ and\ \citenamefont
  {Ecke}(2012)}]{boffetta2012two}%
  \BibitemOpen
  \bibfield  {author} {\bibinfo {author} {\bibfnamefont {G.}~\bibnamefont
  {Boffetta}}\ and\ \bibinfo {author} {\bibfnamefont {R.~E.}\ \bibnamefont
  {Ecke}},\ }\bibfield  {title} {\bibinfo {title} {Two-dimensional
  turbulence},\ }\href@noop {} {\bibfield  {journal} {\bibinfo  {journal}
  {Annual review of fluid mechanics}\ }\textbf {\bibinfo {volume} {44}},\
  \bibinfo {pages} {427} (\bibinfo {year} {2012})}\BibitemShut {NoStop}%
\bibitem [{\citenamefont {Herbert}\ \emph {et~al.}(2012)\citenamefont
  {Herbert}, \citenamefont {Dubrulle}, \citenamefont {Chavanis},\ and\
  \citenamefont {Paillard}}]{herbert2012statistical}%
  \BibitemOpen
  \bibfield  {author} {\bibinfo {author} {\bibfnamefont {C.}~\bibnamefont
  {Herbert}}, \bibinfo {author} {\bibfnamefont {B.}~\bibnamefont {Dubrulle}},
  \bibinfo {author} {\bibfnamefont {P.-H.}\ \bibnamefont {Chavanis}},\ and\
  \bibinfo {author} {\bibfnamefont {D.}~\bibnamefont {Paillard}},\ }\bibfield
  {title} {\bibinfo {title} {Statistical mechanics of quasi-geostrophic flows
  on a rotating sphere},\ }\href@noop {} {\bibfield  {journal} {\bibinfo
  {journal} {Journal of Statistical Mechanics: Theory and Experiment}\ }\textbf
  {\bibinfo {volume} {2012}},\ \bibinfo {pages} {P05023} (\bibinfo {year}
  {2012})}\BibitemShut {NoStop}%
\bibitem [{\citenamefont {Dubkov}\ \emph {et~al.}(2009)\citenamefont {Dubkov},
  \citenamefont {H{\"a}nggi},\ and\ \citenamefont {Goychuk}}]{dubkov2009non}%
  \BibitemOpen
  \bibfield  {author} {\bibinfo {author} {\bibfnamefont {A.}~\bibnamefont
  {Dubkov}}, \bibinfo {author} {\bibfnamefont {P.}~\bibnamefont {H{\"a}nggi}},\
  and\ \bibinfo {author} {\bibfnamefont {I.}~\bibnamefont {Goychuk}},\
  }\bibfield  {title} {\bibinfo {title} {Non-linear {Brownian} motion: the
  problem of obtaining the thermal {Langevin} equation for a non-gaussian
  bath},\ }\href@noop {} {\bibfield  {journal} {\bibinfo  {journal} {Journal of
  Statistical Mechanics: Theory and Experiment}\ }\textbf {\bibinfo {volume}
  {2009}},\ \bibinfo {pages} {P01034} (\bibinfo {year} {2009})}\BibitemShut
  {NoStop}%
\bibitem [{\citenamefont {Goel}\ \emph {et~al.}(1971)\citenamefont {Goel},
  \citenamefont {Maitra},\ and\ \citenamefont {Montroll}}]{goel1971volterra}%
  \BibitemOpen
  \bibfield  {author} {\bibinfo {author} {\bibfnamefont {N.~S.}\ \bibnamefont
  {Goel}}, \bibinfo {author} {\bibfnamefont {S.~C.}\ \bibnamefont {Maitra}},\
  and\ \bibinfo {author} {\bibfnamefont {E.~W.}\ \bibnamefont {Montroll}},\
  }\bibfield  {title} {\bibinfo {title} {On the {Volterra} and other nonlinear
  models of interacting populations},\ }\href@noop {} {\bibfield  {journal}
  {\bibinfo  {journal} {Reviews of modern physics}\ }\textbf {\bibinfo {volume}
  {43}},\ \bibinfo {pages} {231} (\bibinfo {year} {1971})}\BibitemShut
  {NoStop}%
\bibitem [{\citenamefont {Altieri}\ \emph {et~al.}(2021)\citenamefont
  {Altieri}, \citenamefont {Roy}, \citenamefont {Cammarota},\ and\
  \citenamefont {Biroli}}]{altieri2021properties}%
  \BibitemOpen
  \bibfield  {author} {\bibinfo {author} {\bibfnamefont {A.}~\bibnamefont
  {Altieri}}, \bibinfo {author} {\bibfnamefont {F.}~\bibnamefont {Roy}},
  \bibinfo {author} {\bibfnamefont {C.}~\bibnamefont {Cammarota}},\ and\
  \bibinfo {author} {\bibfnamefont {G.}~\bibnamefont {Biroli}},\ }\bibfield
  {title} {\bibinfo {title} {Properties of equilibria and glassy phases of the
  random {Lotka}-{Volterra} model with demographic noise},\ }\href@noop {}
  {\bibfield  {journal} {\bibinfo  {journal} {Physical Review Letters}\
  }\textbf {\bibinfo {volume} {126}},\ \bibinfo {pages} {258301} (\bibinfo
  {year} {2021})}\BibitemShut {NoStop}%
\bibitem [{\citenamefont {Suweis}\ \emph {et~al.}(2024)\citenamefont {Suweis},
  \citenamefont {Ferraro}, \citenamefont {Grilletta}, \citenamefont {Azaele},\
  and\ \citenamefont {Maritan}}]{suweis2024generalized}%
  \BibitemOpen
  \bibfield  {author} {\bibinfo {author} {\bibfnamefont {S.}~\bibnamefont
  {Suweis}}, \bibinfo {author} {\bibfnamefont {F.}~\bibnamefont {Ferraro}},
  \bibinfo {author} {\bibfnamefont {C.}~\bibnamefont {Grilletta}}, \bibinfo
  {author} {\bibfnamefont {S.}~\bibnamefont {Azaele}},\ and\ \bibinfo {author}
  {\bibfnamefont {A.}~\bibnamefont {Maritan}},\ }\bibfield  {title} {\bibinfo
  {title} {Generalized {Lotka}-{Volterra} systems with time correlated
  stochastic interactions},\ }\href@noop {} {\bibfield  {journal} {\bibinfo
  {journal} {Physical Review Letters}\ }\textbf {\bibinfo {volume} {133}},\
  \bibinfo {pages} {167101} (\bibinfo {year} {2024})}\BibitemShut {NoStop}%
\end{thebibliography}%

\newpage

\section*{End Matter}
\appendix
\section{Details on the numerical simulations}
To simulate a system of interacting LV dynamics of the form~\eqref{eq:det-model-sigma}, inspired by~\cite{goel1971volterra}, we consider the following model:
\begin{subequations}
\label{eq:dynsym}
\begin{eqnarray}
   \dot{q}&=&e^p-1+\lambda\sqrt{\gamma}\,\frac{\sum_{j\in B} (e^{p_j-1})}{\sqrt{|B|}} \\
   \dot{p}&=&1-e^q+\lambda\sqrt{\gamma}\,\frac{\sum_{j\in B} (1- e^{q_j})}{\sqrt{|B|}} \\
   \dot{q}_i&=& \lambda^2 \cbr{e^{p_i}-1+\frac{\sum_{j\in B_i} (e^{p_j}-1)}{\sqrt{|B_i|}}}\label{eq:bathq}\\
   &+&\lambda\sqrt{\gamma} \chi_B(i)\frac{e^p-1}{\sqrt{|B|}}\nonumber\\
   \dot{p}_i&=&\lambda^2 \cbr{1-e^{q_i}+\frac{\sum_{j\in B_i} (1- e^{q_j})}{\sqrt{|B_i|}}} \label{eq:bathp}\\
   &+&\lambda\sqrt{\gamma}\chi_B(i)\frac{1-e^q}{\sqrt{|B|}}.\nonumber
\end{eqnarray}
\end{subequations}
Here $(q,p)$ represents the slow degree of freedom, while $(q_i,p_i)$, with $i=1,\dots, N$, are the fast variables. The slow dynamics interacts with some of the fast variables, extracted randomly: we denote by $B$ the set of their indices, $|B|$ the number of its elements and $\chi_B(i)$ its characteristic function, which is equal to 1 if $i\in B$, zero otherwise. Similarly, each particle $i$ of the bath reciprocally interacts with other fast variables, extracted randomly, whose set we denote by $B_i$. In our simulations we take $N=256$, $|B|=16$, $|B_i|=4$ for all $i$, and we ensure that the graph established by the interactions is connected. We always consider the interacting case $\gamma=1$.

Let us now check the validity of Eq.~\eqref{eq:div-free-ms}. In the notation of Eq.~\eqref{eq:det-model-sigma},
$$
f(\vect{x})=\begin{pmatrix}
    e^p-1\\
    1- e^{q}
\end{pmatrix}\quad \Rightarrow \quad \nabla_x f(\vect{x})=0\,.
$$
Similarly, $f_b(\vect{y})$ is a 2N-dimensional vector whose $i$th and $(i+1)$th components are given by the terms proportional to $\lambda^2$ on the rhs of Eqs.~\eqref{eq:bathq} and~\eqref{eq:bathp}. The odd components do not depend on the $\{q_i\}$, while the even ones do not depend on the $\{p_i\}$: one has therefore $\nabla_y f_b(\vect{y})=0$. The same reasoning shows that also the interactions $g(\vect{x})$ and $g_b(\vect{y})$, whose components are the terms proportional to $\lambda \sqrt{\gamma}$ on the rhs of Eq.~\eqref{eq:dynsym}, have vanishing gradients, hence verifying Eq.~\eqref{eq:div-free-ms}.

Let us now focus on the conserved quantities. We define
\begin{equation}
    H(\vect{x})=e^q-q+e^p-p-2
\end{equation}
and
\begin{equation}
     H_b(\vect{y})=\sum_{i=1}^N\cbr{e^{q_i}-{q_i}+e^{p_i}-{p_i}-2}\,.
\end{equation}
It is easy to verify that Eq.~\eqref{eq:en_pres_a}  is always fulfilled. We now define the total energy as
$$
H_{\text{tot}}(\vect{x},\vect{y})=H(\vect{x})+H_b(\vect{y})\,.
$$
Let us notice that $H_{\text{tot}}(\vect{x},\vect{y})$ is \textit{not} the Hamiltonian of the total system, which indeed does not admit a symplectic structure. This quantity happens to fulfill Eq.~\eqref{eq:en_pres_b}: this is true for our model, but it is not obvious in general, because the interactions may spoil such conservation law in the $\gamma>0$ regime. In that case, the conserved $H_{\text{tot}}$, if it exists, may include an interaction term. Finally, we observe that $\op(q,p)=(p,q)$ and $\op_b({(q_i,p_i)})=\op_b({(p_i,q_i)})$ are TR operators for the uncoupled dynamics and for the interactions.

We initialize the system in such a way that the self-energy of each particle,
$$
E_i=e^{q_i}-q_i+e^{p_i}-p_i-2\,,
$$
is equal to 1. We then implement the dynamics~\eqref{eq:dynsym} using a symplectic algorithm (Velocity Verlet Update), with time step $\Delta t=0.02/\lambda^2$. Adopting a symplectic algorithm to simulate non-Hamiltonian dynamics may appear counterintuitive: actually, since the energy function is separable in the $q$'s and $p$'s, this class of integration schemes leads to stable energy conservation even in this case. We let the system evolve until the empirical distribution of the slow degree of freedom reaches stationarity (typically, a few hundreds time units), discard the initial transient dynamics and start measuring from there the relevant observables. We run simulations for more than $3\cdot 10^5$ periods of the slow dynamics. By fitting the stationary pdf of the slow particle, we empirically find $\beta \simeq 1.13$. By applying Eq.~\eqref{eq:pert-diffusion} we compute $D_{qq}\simeq D_{pp} \simeq 0.77$, $D_{qp}\simeq D_{pq} \simeq 0$\,.

%\printbibliography 
\end{document}

\title{Supplemental Material}

%\title{Reversible stochastic processes in the absence of parity rules}
\title{Thermal equilibrium with generalized time-reversal symmetry: Supplemental Material}
\author{Dario Lucente}
\affiliation{Department of Mathematics \& Physics, University of Campania “Luigi Vanvitelli”, Viale Lincoln 5, 81100 Caserta, Italy}
\author{Marco Baldovin} \email{marco.baldovin@cnr.it}
\affiliation{Institute for Complex Systems, CNR, Universit\`a Sapienza, I-00185, Rome, Italy}
\author{Massimiliano Viale}
\affiliation{Institute for Complex Systems, CNR, Universit\`a Sapienza, I-00185, Rome, Italy}
\author{Angelo Vulpiani}
\affiliation{Institute for Complex Systems, CNR, Universit\`a Sapienza, I-00185, Rome, Italy}
\affiliation{Department of Physics, University of Rome “La Sapienza”, P.le Aldo Moro 5, 00185 Rome, Italy}
\date{\today}

\renewcommand{\theequation}{S.\arabic{equation}}
\renewcommand{\thefigure}{S.\arabic{figure}} 

\begin{abstract}
     The Supplemental Material presented here includes: a discussion about reversibility in deterministic systems (Sec.~\ref{sec:det}), the derivation of Eqs.~(7),~(9) (Sec.~\ref{sec:stoch}) and~(15) (Sec.~\ref{sec:pavliotis}), and a series of examples with non-trivial time-reversal parity rules (Sec.~\ref{sec:example}).
\end{abstract}

\maketitle
\section{Deterministic reversibility}
\label{sec:det}
%We start by recalling the property that defines reversibility in deterministic systems~\cite{lamb1998time}. 
% Consider a deterministic system $\vect{x}\in\mathbb{R}^n$ evolving as
Consider a variable \new{$\vect{x}$ evolving in the phase space $\Omega$ according to the deterministic rule}
\begin{equation}
\label{eq:dyn}
\frac{d \vect{x}}{d t} = f\lx(\vect{x}\rx)\,.
\end{equation}
We will write
\begin{equation}
\vect{x}(t) =\varphi^t(\vect x(0))\,, 
\end{equation}
where $\varphi^t$ is the semigroup corresponding to the time evolution. Although this is not needed in order to have reversibility in dynamical systems~\cite{lamb1998time}, we require that $f$ 
%has the properties
%\begin{align}
%    &\nabla \cdot F(\vect{x})=0\,,\label{eq:divergence_free}	\\
%    &F(\vect{x})\cdot \nabla H(\vect{x})=0\,,\label{eq:orthogonal}
%\end{align}
%where $H(\vect{x})$ represent a conserved quantity along the dynamics, which we will refer to as the energy of the system. 
%These conditions ensure that the dynamics is constrained on a constant energy surface $H(\vect{x})=E$  and, moreover, that the evolution preserves the volume of the phase space $\Omega$ ($S^t$ is a unitary operator).
\new{
verifies
\begin{equation}
    \nabla \cdot f(\vect{x})=0\,,\label{eq:divergence_free}
\end{equation}
ensuring that the evolution preserves the volume in $\Omega$ ($\varphi^t$ is a unitary operator). Moreover, we require the existence of a smooth function $H: \Omega \to \mathbb{R}$, which we will refer to as the energy of the system, satisfying
\begin{equation}
    f(\vect{x})\cdot \nabla H(\vect{x})=0\,.\label{eq:orthogonal}
\end{equation}
The energy is therefore conserved along the motion, and the dynamics is constrained on a constant energy surface $H(\vect{x})=E$.
In particular, if 
$$
\vect{x}=(\vect{q},\vect{p})\in \mathbb{R}^{2n}\,,\quad f\lx(\vect{x}\rx)=J\nabla H(\vect{x})\,, \quad J=\begin{pmatrix}0 & I_{n} \\ -I_{n} & 0 \end{pmatrix}\,,
$$
where $I_n$ is the $n\times n$ identity matrix, then the dynamics is symplectic, and  $H(\vect{x})$ is the Hamiltonian. However, we will keep the discussion at a more general level, and we will not assume a symplectic structure.}

%If the dynamics is Hamiltonian, the conserved quantity is clearly the Hamiltonian $H(\vect{x})$ itself, and $i\mathcal{L}=-\{H, \cdot\}$, where we used braces to represent Poisson's brackets. 

The dynamics~\eqref{eq:dyn} is \textit{reversible} if an involution $\op$ (i.e. an operator $\op : \Omega \to \Omega$ such that $\op\circ\op$ is the identity) exists, having the following properties:
\begin{enumerate}
\item $\op$ is a symmetry of the energy $H(\vect{x})$:
    \begin{equation}
    \label{eq:prop1}
        H(\op(\vect{x}))=H(\vect{x})\,;
    \end{equation}
\item it verifies
    \begin{equation}
    \label{eq:prop2a}
    \op \circ \varphi^t = \varphi^{-t} \circ \op\,,
    \end{equation}
or, equivalently,
    \begin{equation}
    \label{eq:prop2b}
    M f(\vect{x})=-f(\op(\vect{x})) 
    \end{equation}
with 
\begin{equation}
    \label{eq:defM}
M_{ij}(\vect{x})\equiv\frac{\partial \op_i (\vect{x})}{\partial x_j}\,.
    \end{equation}
\end{enumerate}
Properties~\eqref{eq:dyn} and~\eqref{eq:prop2b}  lead to 
\begin{equation}
    \label{eq:revdyn}
-\frac{d (\op (\vect{x}))}{dt}= f (\op( \vect{x}))\,,
\end{equation}
which is formally identical to \eqref{eq:dyn} in the variables $\op (\vect{x})$, but for a minus sign in front of the time. \new{In this sense, the involution $\varepsilon$ \textit{reverts} the dynamics.}

The probability distribution  $\cP(\vect{x},t)$ \new{associated to the deterministic system~\eqref{eq:dyn}}  evolves according to the Liouville equation
\begin{equation}
\label{eq:Liouville}
\partial_t \cP(\vect{x},t) + \nabla\cdot\lx[f(\vect{x}) \cP(\vect{x},t)\rx]=\partial_t \cP(\vect{x},t) + f(\vect{x})\cdot\nabla\cP(\vect{x},t)=0\,,
\end{equation}
where we have used Eq.~\eqref{eq:divergence_free}, 
and condition~\eqref{eq:orthogonal} guarantees that 
\begin{equation}
    \label{eq:fH}
    \cP_s(\vect{x})=F(H(\vect{x}))
\end{equation}
is a stationary solution for every function $F$ ensuring normalization. 
Reversibility condition~\eqref{eq:prop2b} is strictly related to the possibility of identifying detailed balance symmetry for probability distributions evolving through Eq.~\eqref{eq:Liouville}. 
Generally speaking, detailed balance holds if 
 \begin{equation}
\label{eq:detbal}
   \cP_s(\vect{x}')W_t\cbr{\vect{x} | \vect{x}'}=\cP_s(\op (\vect{x}))W_t\cbr{ \op (\vect{x}') | \op(\vect{x}) }
\end{equation}
being $W_t\cbr{\vect{x} | \vect{x}'}$ the propagator of the dynamics. %For deterministic system, $W_t\cbr{\vect{x} | \vect{x}'}=\delta(\vect{x}-S^t \vect{x}')$ and the existence of a reversing symmetry $\op$ implies  
For deterministic systems,  $W_t\cbr{\vect{x} | \vect{x}'}=\delta(\vect{x}-\varphi^t (\vect{x}'))$ and the existence of a reversing symmetry $\op$ implies
\begin{equation}
\label{eq:detbal_det}
%\begin{aligned}
W_t( \op (\vect{x}') | \op (\vect{x}) )=\delta\sbr{\op(\vect{x}')-\varphi^{t}(\op (\vect{x}))}
=\delta\sbr{\varphi^{-t}(\op(\vect{x}'))-\op(\vect{x})}=\delta\sbr{\op( \varphi^{t}(\vect{x}'))-\op( \vect{x})}
=W_t( \vect{x} |  \vect{x}' )\,,
%\end{aligned}
\end{equation}
where we used that both $\varphi^{t}$ and $\op$ are invertible and have unitary Jacobian. Therefore, Eq.~\ref{eq:detbal} boils down to $\cP_s(\vect{x})=\cP_s(\op\cbr{\vect{x}})$, which
%, for symplectic dynamics,
is guaranteed by condition~\eqref{eq:prop1} \new{if condition~\eqref{eq:fH} holds}.

\section{Stochastic reversibility}
\label{sec:stoch}

\subsection{Detailed-Balance for generic stationary states}

Let us now consider the stochastic (It\={o}) differential equation
\begin{equation}
\label{eq:dyn_stoch}
\frac{d \vect{x}}{d t} = A\lx(\vect{x}\rx)   +\vect{\xi}\,, 
\end{equation}
with $\av{\vect{\xi}(t)\vect{\xi}^T(t')}=2\gamma D(\vect{x})\delta(t-t')$, $D=D^T$ and $\gamma>0$.
\new{ We want to find necessary conditions for the detailed balance condition~\eqref{eq:detbal} to hold.}
The propagator $W_t(\vect{x}|\vect{x}')$ satisfies the forward and backward Kolmogorov equations 
\begin{subequations}
\begin{equation}
\label{eq:fw_generic}
    %\partial_t W_{X_1\to X_2}(t) = -\nabla_{X_2} \sbr{(J-\beta D)(\nabla_{X_2}H(\vect X_2))W_{X_1\to X_2}(t)-D \nabla_{X_2}W_{X_1\to X_2}(t)}
    \partial_t W_t(\vect{x}|\vect{x}') = -\sum_i \partial_i \lx[A_i(\vect{x})W_t(\vect{x}|\vect{x}')  - \gamma \sum_{j} \partial_j  D_{ij}(\vect{x}) W_t(\vect{x}|\vect{x}') \rx]=\mathcal{L}_{\vect{x}}^{\it{fw}}\sbr{W_t\cbr{ \vect{x} | \vect{x}'} }
%\partial_t W(X'|X) &= \sum_i A(X)_i \partial_i W_t(X'|X) + \frac{1}{2} \sum_{ij} B(X)_{ij}\partial_i\partial_j W_t(X'|X) = \cD_r(X) \circ W_t(X'|X) \nonumber
\end{equation}
\begin{equation}
\label{eq:backw_generic}
    %\partial_t W_{X_1\to X_2}(t) = -\nabla_{X_2} \sbr{(J-\beta D)(\nabla_{X_2}H(\vect X_2))W_{X_1\to X_2}(t)-D \nabla_{X_2}W_{X_1\to X_2}(t)}
    \partial_t W_t(\vect{x}|\vect{x}') = \sum_i \lx[A_i(\vect{x}')\partial_i' W_t(\vect{x}|\vect{x}')  + \gamma \sum_{j}D_{ij}(\vect{x}')  \partial_i' \partial_j' W_t(\vect{x}|\vect{x}') \rx]=\mathcal{L}_{\vect{x}'}^{\it{bw}}\sbr{W_t\cbr{ \vect{x} | \vect{x}'} }\,,
%\partial_t W(X'|X) &= \sum_i A(X)_i \partial_i W_t(X'|X) + \frac{1}{2} \sum_{ij} B(X)_{ij}\partial_i\partial_j W_t(X'|X) = \cD_r(X) \circ W_t(X'|X) \nonumber
\end{equation}
\end{subequations}
where we used the shorthand notation
$$
\partial_i\equiv\frac{\partial}{\partial x_i}\,,\quad \quad\partial'_i\equiv\frac{\partial}{\partial x'_i}\,,
$$
while the stationary distribution $\cP_S(\vect{x})$ verifies
\begin{equation}
\label{eq:fp_generic}
\mathcal{L}_{\vect{x}}^{\it{fw}}\sbr{\cP_S(\vect{x})}=\mathcal{L}_{\vect{x}}^{\it{bw}}\sbr{\cP_S(\vect{x})}=0\,.
    %\partial_t W_{X_1\to X_2}(t) = -\nabla_{X_2} \sbr{(J-\beta D)(\nabla_{X_2}H(\vect X_2))W_{X_1\to X_2}(t)-D \nabla_{X_2}W_{X_1\to X_2}(t)}
    %\sum_i \partial_i \lx[A_i(\vect{x})\cP_S(\vect{x})  - \sum_{j} \partial_j ( D_{ij}(\vect{x}) \cP_S(\vect{x}) )\rx]=0
%\partial_t W(X'|X) &= \sum_i A(X)_i \partial_i W_t(X'|X) + \frac{1}{2} \sum_{ij} B(X)_{ij}\partial_i\partial_j W_t(X'|X) = \cD_r(X) \circ W_t(X'|X) \nonumber
\end{equation}
By plugging the detailed balance condition~\eqref{eq:detbal} for a given involution $\op$, i.e.
$$
W_t(\vect{x}|\vect{x}')= \frac{\cP_S(\op(\vect{x}))}{\cP_S(\vect{x}')}W_t(\op(\vect{x}')|\op(\vect{x}))\,,
$$ into Eq.~\eqref{eq:fw_generic}, we obtain 
\begin{equation} 
\frac{\cP_S(\op (\vect{x}))}{\cP_S(\vect{x}')}\partial_t W_t(\op(\vect{x}')|\op(\vect{x})) = \frac{1}{\cP_S(\vect{x}')}\mathcal{L}_{\vect{x}}^{\it{fw}}\sbr{\cP_S(\op (\vect{x}))W_t(\op(\vect{x}')|\op(\vect{x}))}\,.
\end{equation}
Since $W_t(\op(\vect{x}')|\op(\vect{x}))$ evolves according to Eq.~\eqref{eq:backw_generic}, we arrive at \new{the condition}
\begin{equation} 
\label{eq:op-equiv}
\cP_S(\op( \vect{x}))\mathcal{L}_{\op( \vect{x})}^{\it{bw}}\sbr{W_t(\op(\vect{x}')|\op(\vect{x}))} =\mathcal{L}_{\vect{x}}^{\it{fw}}\sbr{\cP_S(\op (\vect{x}))W_t(\op(\vect{x}')|\op(\vect{x}))}\,.
\end{equation}
Equation~\eqref{eq:op-equiv} can be equivalently rewritten, with the change of variables $(\op (\vect{x}'),\op (\vect{x}))\to (\vect{x}_1,  \vect{x}_2)$, as 
$$
\cP_S(\vect{x}_2)\mathcal{L}_{\vect{x}_2}^{\it{bw}}\sbr{W_t(\vect{x}_1|\vect{x}_2)} =\mathcal{L}_{\op (\vect{x}_2)}^{\it{fw}}\sbr{\cP_S( \vect{x}_2)W_t(\vect{x}_1|\vect{x}_2)}\,,
$$
which is Eq.~(8) of the main text.
From now on, for the sake of brevity, we use the following shorthand symbols:
$$
    \hpartial_i \equiv \frac{\partial}{\partial{\op_i (\vect{x})}}\,,\quad \quad
    \widehat{W}_t \equiv W_t(\op (\vect{x}')|\op (\vect{x}))\,,\quad \quad
    \widehat{g} \equiv g(\op (\vect{x}))\,,
$$
where $g$ is a generic observable.
The r.h.s. of Eq.~\eqref{eq:op-equiv} consists of two terms: one of them involves the drift $A$, the other one is related to the diffusion matrix $D$.  The former can be written as
\begin{equation}
\label{eq:op-equiv-drift}
-\sum_i \partial_i A_i \widehat{\cP}_S \widehat{W}_t =-\sum_i \lx[\lx(\partial_i A_i \widehat{\cP}_S\rx)\widehat{W}_t
+A_i\widehat{\cP}_S \partial_i\widehat{W}_t \rx]\,,
\end{equation}
while the latter reads
\begin{equation}
\label{eq:op-equiv-diff}
\begin{aligned}
\gamma \sum_{ij} \partial_i \partial_j  D_{ij} \widehat{\cP}_S\widehat{W}_t &=\gamma \sum_{ij} \partial_i \lx[ (\partial_j  D_{ij} \widehat{\cP}_S)\widehat{W}_t + D_{ij} \widehat{\cP}_S\partial_j\widehat{W}_t\rx]=\\
&=\gamma \sum_{ij} \lx[ \partial_i  (\partial_j  D_{ij} \widehat{\cP}_S) \widehat{W}_t+\lx(\partial_iD_{ij} \widehat{\cP}_S\rx)\partial_j\widehat{W}_t+D_{ij}\widehat{\cP}_S\partial_i\partial_j\widehat{W}_t\rx]=\\
&=\gamma \sum_{ij}\sbr{\lx( \partial_i \partial_j  D_{ij} \widehat{\cP}_S \rx) \widehat{W}_t+2\lx(\partial_j  D_{ij} \widehat{\cP}_S\rx)\partial_i\widehat{W}_t+ D_{ij}\widehat{\cP}_S\partial_i\partial_j\widehat{W}_t}\,.%=\\
%&= \sum_{ij}\lx( \partial_i \partial_j  D_{ij} \widehat{\cP}_S \rx) \widehat{W}_t+\sum_{ij} D_{ij}\widehat{\cP}_S\partial_i\partial_j\widehat{W}_t+2\sum_i \widehat{\cP}_S \Gamma_i \partial_i\widehat{W}_t
\end{aligned}
\end{equation}
\new{Requiring detailed balance to hold implies that} the terms proportional to $\widehat{W}_t$ in Eqs~\eqref{eq:op-equiv-drift}-\eqref{eq:op-equiv-diff} vanish, because they are not present in the l.h.s. of Eq.~\eqref{eq:op-equiv}, which depends only on derivatives of $\widehat{W}_t$. Therefore one has
\begin{equation}
   \widehat{W}_t\sum_i\partial_i\lx[-A_i\widehat{\cP}_S+\gamma \sum_j\partial_j  D_{ij} \widehat{\cP}_S\rx]=0  \implies
   \widehat{W}_t \mathcal{L}_{\vect{x}}^{\it{fw}} \widehat{\cP}_S =0  
\end{equation}
hence
\begin{equation}
    \widehat{\cP}_S\equiv \cP_S(\op ( \vect{x}))=\cP_S( \vect{x})\,.
\end{equation}
Upon imposing the previous condition and defining
\begin{equation}
\label{eq:gammatilde}
\widetilde{\Gamma}_i (\vect{x})\equiv \frac{1}{P_S(\vect{x})}\sum_j\partial_j  D_{ij}(\vect{x}) \cP_S(\vect{x})\,,
\end{equation}
Eq.~\eqref{eq:op-equiv} can be rewritten as
\begin{equation}
\label{eq:detailed_balance_fokker_planck_general}
\mathcal{L}_{\op( \vect{x})}^{\it{bw}}\sbr{W_t(\op(\vect{x}')|\op(\vect{x}))}= \sum_i\lx[ \lx(2\gamma \widetilde{\Gamma}_i(\vect{x})-A_i(\vect{x})\rx)\partial_iW_t(\op(\vect{x}')|\op(\vect{x}))+\gamma \sum_j D_{ij}\partial_i\partial_jW_t(\op(\vect{x}')|\op(\vect{x}))\rx]\,.
\end{equation}
Recalling definition~\eqref{eq:defM}, it is immediate to verify that
\begin{equation}
\label{eq:trasfM}
	\partial_i g(\vect{x})  =\sum_j M^T_{ij}\hpartial_j g (\vect{x})\,,
\end{equation}
for every function $g$. The two terms on the r.h.s. of Eq.~\eqref{eq:detailed_balance_fokker_planck_general} read therefore
\begin{align}
   &\sum_i\lx(2 \gamma \widetilde{\Gamma}_i(\vect{x})-A_i(\vect{x})\rx)\partial_i\widehat{W}_t=\sum_{ik}\lx(2 \gamma \widetilde{\Gamma}_i(\vect{x})-A_i(\vect{x})\rx)M_{ik}^T(\vect{x})\hpartial_k\widehat{W}_t\,,\\
&\gamma \sum_{ij} D_{ij}\partial_i\partial_j\widehat{W}_t=\gamma \sum_{ijk} D_{ij}\partial_iM_{jk}^T(\vect{x})\hpartial_k\widehat{W}_t=\gamma \sum_{ijk} D_{ij}\lx(\partial_iM_{jk}^T(\vect{x})\rx)\hpartial_k\widehat{W}_t+\gamma \sum_{ijkl} D_{ij}M_{li}(\vect{x}) M_{jk}^T(\vect{x})\hpartial_l\hpartial_k\widehat{W}_t\,.
\end{align}
The condition for detailed balance  is then obtained from~\eqref{eq:detailed_balance_fokker_planck_general} by imposing the equivalence of drift and diffusion
\begin{align}
\label{eq:noise_condition_general}
D(\op (\vect{x}))&=M(\vect{x})D(\vect{x})M^T(\vect{x})\\
\label{eq:drift_condition_general}
A_k(\op (\vect{x}))&=\sum_i \lx[M_{ki}(\vect{x})\lx(2 \gamma \widetilde{\Gamma}_i(\vect{x})-A_i(\vect{x})\rx)+\gamma \sum_j D_{ij}\partial_iM_{jk}^T(\vect{x})\rx]\,.
\end{align}
Taking into account the explicit expression of $\widetilde{\Gamma}$ and multiplying Eq.~\eqref{eq:drift_condition_general} by $M^{-1}$ we get
\begin{equation}
\label{eq:drift_condition_general_final}
 \sum_k M_{lk}(\op (\vect{x})) A_k (\op (\vect{x}))=-A_l(\vect{x})+2\gamma \cP_S^{-1}(\vect{x})\sum_k \partial_k D_{lk}(\vect{x})\cP_S(\vect{x})+\gamma \sum_{ijk}M^{-1}_{lk}(\vect{x})D_{ij}(\vect{x})\partial_iM_{jk}^T(\vect{x})
\end{equation}
where on the l.h.s. of Eq.~\eqref{eq:drift_condition_general_final}, we have used that, since $\op$ is an involution, $M(\vect{x})=M^{-1}(\op (\vect{x}) )$. 

Let us notice that, in many cases of physical interest, the last term on the r.h.s. of Eq.~\eqref{eq:drift_condition_general_final} identically vanishes. In particular, this happens in the presence of linear involutions $\op$, where $M$ does not depend on $\vect{x}$. Examples are: the usual parity rules flipping the sign of the velocities,  the exchange operator discussed in the symmetric Lotka-Volterra and in point-vortexes model,  the parity for truncated hydrodynamics.

\subsection{Conditions for equilibrium: Eqs.~(7) and~(9)}
\label{sec:cond}

\new{Let us specialize to the case in which} the stochastic dynamics in Eq.~\eqref{eq:dyn_stoch} represents a reversible deterministic system coupled to a thermal bath. The drift can be  divided into two contributions: the deterministic force \new{$f(\vect{x})$ defined by Eq.~\eqref{eq:dyn}} and the dissipation \new{$\gamma \Gamma(\vect{x})$ due to the coupling with the bath}:
\begin{equation}
    A(\vect x)=f(\vect{x})+\gamma \Gamma(\vect{x})\,.
\end{equation}
%. The Langevin equation reads
%\begin{equation}
%    \dot{\vect{x}}=f(\vect{x})+\gamma \Gamma(\vect{x})+\boldsymbol{\xi}
%\end{equation}
%with $\ave{\boldsymbol{\xi}}=0$ and $\ave{\boldsymbol{\xi}(t)\boldsymbol{\xi}^T(t')}=2\gamma D(\vect{x})\delta(t-t')$.
This is exactly the case considered in Eq.~(4) of the main text. 
The corresponding Fokker-Planck equation is
\begin{equation}
    \partial_t \cP(\vect{x},t)=-\nabla\cdot\mathcal{J}(\vect{x},t)
\end{equation}
with
\begin{equation}
    \mathcal{J}_i(\vect{x},t)=\lx[f_i(\vect{x})+\gamma \Gamma_i(\vect{x})\rx] \cP(\vect{x},t)-\gamma \sum_j \partial_{j} D_{ij} (\vect{x})\cP(\vect{x},t)\,.
\end{equation}
\new{We now require} the stationary distribution $\cP_S(\vect{x})$ to be the Boltzmann one, i.e. $\cP_S(\vect{x})=e^{-\beta H}/Z_{\beta}$, where $\beta$ is the inverse temperature and $Z_{\beta}$ is a normalizing factor. We get the stationary condition
\begin{equation}
\label{eq:stationary_measure}
   \nabla\cdot\mathcal{J}^{(S)}(\vect{x})=0
\end{equation}
where
\begin{equation}
    \mathcal{J}_i^{(S)}(\vect{x})=\frac{e^{-\beta H}}{Z_{\beta}}\lx[f_i(\vect{x})+\gamma \Gamma_i(\vect{x})-\gamma e^{\beta H}\sum_j \partial_{j} D_{ij} (\vect{x})e^{-\beta H}\rx]\,. 
\end{equation}

By combining Eq.~\eqref{eq:divergence_free}, Eq.~\eqref{eq:orthogonal} and Eq.~\eqref{eq:stationary_measure}, we obtain
\begin{equation}
   \nabla\cdot\mathcal{J}^{(S)}(\vect{x})=\frac{\gamma}{Z_{\beta}}\sum_{i}\partial_i e^{-\beta H}\lx[\Gamma_i(\vect{x})-e^{\beta H}\sum_j \partial_{j} D_{ij} (\vect{x})e^{-\beta H}\rx]=0  
\end{equation}
whose general solution is 
\begin{equation}
\label{eq:dissipation}
\begin{aligned}    
   \Gamma_i(\vect{x})&=e^{\beta H}\sum_j \partial_{j} \lx[D_{ij} (\vect{x})e^{-\beta H}\rx] +h(\vect{x})\\
   &=\widetilde{\Gamma}(\vect{x})+h(\vect{x})
\end{aligned}
\end{equation}
where $\widetilde{\Gamma}(\vect{x})$ has been defined in Eq.~\eqref{eq:gammatilde}, and  $h(\vect{x})$ is a generic vector field with the property
\begin{equation}
%\label{eq:dissipation}
   \nabla\cdot h(\vect{x})=\beta h(\vect{x})\cdot \nabla H \,.%\frac{\partial \lx[D_{ij} (\vect{x})\cP(\vect{x},t)\rx]}{\partial X_j}%\lx[J\nabla H \cP(\vect{x},t)+\Gamma \cP(x,t)-D\nabla\cP(x,t)\rx]
\end{equation}
%In order to retrieve the deterministic dynamics in the limit $\gamma\to 0$ we set $h(\vect{x})\equiv 0$ in the following.
The above form for $\Gamma(\vect x)$ guarantees that the Boltzmann distribution is approached for $t\to\infty$. 
In particular, choosing $h(\vect x)=0$ leads to
\begin{equation}
\label{eq:dissipation_final}
   \Gamma_i(\vect{x})=e^{\beta H}\sum_j \partial_{j} D_{ij} (\vect{x})e^{-\beta H}\,.%\frac{\partial \lx[D_{ij} (\vect{x})\cP(\vect{x},t)\rx]}{\partial X_j}%\lx[J\nabla H \cP(\vect{x},t)+\Gamma \cP(x,t)-D\nabla\cP(x,t)\rx]
\end{equation}
This is Eq.~(7) of the main text. We will show in Section~\ref{sec:pavliotis} that the choice $h(\vect x)=0$ has a clear physical motivation.

When $\Gamma(\vect x)$ fulfills Eq.~\eqref{eq:dissipation}, the general detailed balance conditions~\eqref{eq:noise_condition_general}-\eqref{eq:drift_condition_general} read
\begin{align}
\label{eq:noise_condition}
D(\op (\vect{x}))&=M(\vect{x})D(\vect{x})M^T(\vect{x})\\
\label{eq:drift_condition}
f_k(\op ( \vect{x}))+\gamma \widetilde{\Gamma}_k(\op (\vect{x}))+\gamma h_k(\op (\vect{x}))&=\sum_i \lx[M_{ki}(\vect{x})\lx(\gamma \widetilde{\Gamma}_i(\vect{x})-f_i(\vect{x}) -\gamma h_i(\vect{x}) \rx)+\gamma \sum_j D_{ij}(\vect{x})\partial_i M_{jk}^T(\vect{x})\rx]\,.
\end{align}
Recalling Eq.~\eqref{eq:prop2b}, Eq.~\eqref{eq:drift_condition} can be written as
\begin{equation}
\sum_i M_{ki}(\vect{x})\widetilde{\Gamma}_i(\vect{x})= \widetilde{\Gamma}_k(\op ( \vect{x}))- \sum_{ij}D_{ij}( \vect{x})\partial_i M_{jk}^T(\vect{x})+{h_k(\op (\vect{x}))+\sum_i M_{ki}(\vect{x})h_i(\vect{x})}\,.\label{eq:dissipation_time_reversal}\end{equation}
As we will prove shortly, the equation above is fulfilled whenever 
\begin{equation}
\label{eq:badcond2}
    h_k(\op (\vect{x}))=-\sum_i M_{ki}(\vect{x})h_i(\vect{x})\,.
\end{equation}
The above equation is trivially satisfied whenever $h(\vect x)=0$, which is the case in most physics-inspired applications, see Section~\ref{sec:pavliotis} below.
\new{Indeed, rovided that Eq.~\eqref{eq:badcond2} is satisfied, the proof that Eq.~\eqref{eq:noise_condition} yields Eq.~\eqref{eq:dissipation_time_reversal} proceeds by direct computation:}
\begin{equation}
\begin{aligned}
\sum_i M_{ki}\widetilde{\Gamma}_i&=\sum_{ij}M_{ki}e^{\beta H}\partial_jD_{ij}e^{-\beta H}\\
&=\sum_{ijlm}M_{ki}e^{\beta H}\partial_j M_{il}^{-1}\widehat{D}_{lm}M_{mj}^{-T}e^{-\beta H}\\
&=\sum_{ijlm}M_{ki} M_{il}^{-1}M_{mj}^{-T} e^{\beta H}\partial_j \widehat{D}_{lm}e^{-\beta H}
+\sum_{ijlm}M_{ki}\widehat{D}_{lm}\partial_j M_{il}^{-1}M_{mj}^{-T}\\
&=\sum_{jlmn}\delta_{kl}M_{mj}^{-T} e^{\beta H}M_{jn}^{T}\widehat{\partial}_n \widehat{D}_{lm}e^{-\beta H}
+\sum_{ijlm}M_{ki}\widehat{D}_{lm}\partial_j M_{il}^{-1}M_{mj}^{-T}\\
&=\sum_{jmn}M_{mj}^{-T}M_{jn}^{T} e^{\beta H}\hpartial_n \widehat{D}_{km}e^{-\beta H}
+\sum_{ijlm}\widehat{D}_{lm}\partial_j M_{ki} M_{il}^{-1}M_{mj}^{-T}-\sum_{ijlm}M_{il}^{-1}M_{mj}^{-T}\widehat{D}_{lm}\partial_j M_{ki}\\
&=\sum_{n} e^{\beta \widehat{H}}\hpartial_n\widehat{D}_{kn}e^{-\beta \widehat{H}}+\sum_{jm}\widehat{D}_{km}\partial_j M_{mj}^{-T}-\sum_{ij}D_{ij}\partial_j M_{ki}\\
&=\widehat{\widetilde{\Gamma}}_k+\sum_{jm}\widehat{D}_{km}\partial_j \widehat{M}_{jm}-\sum_{ij}D_{ji}\partial_j M_{ik}^T\\
&=\widehat{\widetilde{\Gamma}}_k-\sum_{ij}D_{ij}\partial_i M_{jk}^T\,.
\end{aligned}
\end{equation}
In the last step we used  
\begin{equation}
    \sum_j \partial_j M_{jm}(\op(\vect{x}))=\sum_{jk} M^{-1}_{lj}(\vect{x})\partial_l M_{jm}(\vect{x})=\sum_{jk} M^{-1}_{lj}(\vect{x})\frac{\partial^2 \op_j(\vect{x})}{\partial_{x_l}\partial_{x_m}}=\sum_{jk} M^{-1}_{lj}(\vect{x})\partial_m M_{jl}(\op(\vect{x}))=\partial_m \ln \lx|\mathrm{det} M(\vect{x})\rx|\,.
\end{equation}
The above expression is identically zero because we assume $\op(\vect x)$ to be volume-preserving, i.e. $|\mathrm{det} M(\vect{x})|=1$.

\section{Multiscale dynamics at equilibrium}
\label{sec:pavliotis}

In this section we want to derive an effective diffusion dynamics for a slow system in contact with a thermal bath. As a starting point we consider the multiscale system
\begin{equation}
    \begin{cases}
        \dot{\vect{x}} = f(\vect{x})+\lambda \,g(\vect{x},\vect{y})\,;\\
        \dot{\vect{y}} = \lambda^2 f_b(\vect{y})+\lambda \,g_b(\vect{x},\vect{y})\,.\\
    \end{cases}
    \label{eq:det-model-lambda}
\end{equation}
where  $\lambda$ is a large parameter that enforces a time-scale separation between the system degrees of freedom $\vect x$ and those of the thermal bath $\vect{y}$. The parameter $\gamma\ge0$ appearing in the main text, that represents the strength of the coupling between $\vect{x}$ and $\vect{y}$, has been included in definition of $g$ and $g_b$ for simplicity. %The total energy $H_{\text{tot}}(\vect{x},\vect{y})$ is given by the sum of three energies
Formally, we define a family of total energies
\begin{equation}
    H_{\mathrm{tot}}(\vect x,\vect y) = H(\vect x)+H_b(\vect y)
    +
    H_\lambda(\vect x,\vect y),
    \label{eq:Htot-general}
\end{equation}
where $H(\vect{x})$ and $H_b(\vect{y})$ are the energies of the uncoupled systems while $H_{\lambda}(\vect{x},\vect{y})$ is the interaction term. To be consistent with a thermodynamic description, we assume the existence of a scaling function $N_\lambda$ such that the energy of the bath $H_b(\vect{y})$ and the total energy $H_{tot}=E$ are extensive in $N_\lambda$ , that is
\begin{align}
    &H_b^{(\lambda)}(\vect y) = O(N_\lambda)\,;\\
    &E=O(N_\lambda)\,.
\end{align}
Conversely, we impose the energy of the subsystem $\vect x$, $H(\vect x)$, to be order $O(1)$ and the interaction energy $H_{\lambda}(\vect{x},\vect{y})$ to vanish as some scaling function $r_\lambda$ in the limit $\lambda\to\infty$, i.e.
\begin{align}
    &H(\vect{x}) = O(1)\,;\\
    &H_{\lambda}(\vect{x},\vect{y})=O(r_\lambda)\,;
\end{align}
with $\lim_{\lambda\to\infty} r_\lambda=0$. In addition, as will become clear in Sec.~\ref{sec:derivation-drift-ms}, to enforce a non-vanishing thermodynamic force at order $O(1)$ and to neglect interactions, we also assume that 
\begin{align}
    &\nabla_x H(\vect{x}) = O(1)\,;\\
    &\nabla_x H_{\lambda}(\vect{x},\vect{y})=O(r_\lambda)\,. 
\end{align}
%\begin{equation}
%	H_{\text{tot}}(\vect{x},\vect{y}) = H(\vect{x})+H_b(\vect{y})+H_{\lambda}(\vect{x},\vect{y})
%\end{equation}
 %where $H(\vect{x})$ and $H_b(\vect{y})\gg H(\vect{x})$ are the energies of the uncoupled systems while $H_{\lambda}(\vect{x},\vect{y})$ is the interaction term that we require to vanish in the $\lambda\to\infty$ limit. We require the dynamics of the coupled system ($\gamma>0$) and those of the isolated ones in the decoupled limit ($\gamma=0$)  to fulfill: \\
 For dealing with an equilibrium physical system, we require the coupled dynamics and the isolated ones in the decoupled limit to fulfill: \\
 (i) volume preservation
\begin{equation}
\nabla_x\cdot f=\nabla_y\cdot f_b =0\,,\quad\quad\nabla_x\cdot g=-\nabla_y\cdot g_b\,; \label{eq:div-free-ms}
\end{equation}
(ii) conservation of the energy
\begin{align}
 f \cdot \nabla_x H &= f_b \cdot \nabla_y H_b=0\,,\\
g \cdot \nabla_x H_{\text{tot}} &= -g_b \cdot \nabla_y H_{\text{tot}} \,;
\label{eq:energy-preservation-ms}
\end{align}
(iii) reversibility under $\op_{\text{tot}}(\vect x,\vect y)=(\op(\vect x),\op_b(\vect y))$
\begin{align}
&f(\op (\vect x))=-M(\vect x)f(\vect x)\,,\quad\quad g(\op (\vect x), \op_b (\vect y))=-M(\vect x)g(\vect x, \vect y)\,;\\
&f_b(\op_b (\vect y))=-M_b(\vect y)f_b(\vect y)\,,\quad g_b(\op (\vect x), \op_b (\vect y))=-M_b(\vect y)g_b(\vect x, \vect y)\,.
\label{eq:reversibility-ms}
\end{align}
In the rest of the section we will show that for a system with this properties, under suitable ergodicity assumptions on the thermal bath $\vect y$~\cite{pavliotis2008multiscale,kelly2017deterministic}, the slow system $\vect x$ is well described by a stochastic process whose drift and diffusion satisfy Eqs.~(7) and~(9). In particular, the diffusion equation for observables of the slow variable $\vect{x}$ is derived in Sec.~\ref{sec:derivation-langevin-ms}. In Sec.~\ref{sec:derivation-drift-ms} and Sec.~\ref{sec:derivation-symmetry-diffusion-ms} it is shown that Eqs.~(7) and~(9) hold provided that conditions~\eqref{eq:div-free-ms}-\eqref{eq:energy-preservation-ms} and~\eqref{eq:reversibility-ms} are fulfilled, respectively. 

\subsection{Derivation of Langevin dynamics}\label{sec:derivation-langevin-ms}
To derive the diffusion dynamics for the slow degrees of freedom of the system~\eqref{eq:det-model-lambda}, we should assume suitable ergodicity assumption on the fast flow induced by $f_b(\vect{y})$~\cite{pavliotis2008multiscale,kelly2017deterministic}. 
For simplicity, following~\cite{pavliotis2008multiscale} let us also assume 
\begin{equation}
\mathbb{E}_{\vect{x}}[g(\vect{x},\vect{y})] = 0\label{eq:Fredholm}\,,
\end{equation}
where $\mathbb{E}_{\vect{x}}$ represents the expected value over the stationary distribution $\rho_{\vect{x}}(\vect{y})$ in a dynamics where $\vect{x}$ is fixed.
This condition will be necessary in order to invoke the Fredholm alternative, and it corresponds to the choice $h(\vect x)\equiv 0$. For sake of completeness, let us mention that, since the derivation is done in perturbation theory, the less restrictive conditions 
\begin{equation}
\mathbb{E}_{\vect{x}}[g(\vect{x},\vect{y})] = O(\lambda^{-1}) \equiv \lambda^{-1}\, \bar{g}(\vect{x})\,,\quad\quad \tilde{g}(\vect{x},\vect{y})=\lx[g(\vect{x},\vect{y})- \lambda^{-1}\, \bar{g}(\vect{x})\rx]=O(1)\,;
\end{equation}
would suffice. 

The Liouville equation of the system reads
\begin{equation}
\partial_t \rho_t = -\left(\lambda^2 \mathcal{L}_2^* +\lambda \mathcal{L}_1^* + \mathcal{L}_0^*\right)\rho_t,
\label{eq:liouville_ms}
\end{equation}
where
\begin{align}
\mathcal{L}_2 &= f_b(\vect{y}) \cdot \nabla_y, \\
\mathcal{L}_1 &= g_b(\vect{x},\vect{y}) \cdot \nabla_y + g(\vect{x},\vect{y}) \cdot \nabla_x\,, \\
\mathcal{L}_0 &= f(\vect{x}) \cdot \nabla_x\,,
\end{align}
and the star $^*$ denotes adjoint operators. 

Let us consider a generic observable $\Psi(\vect{x},\vect{y})$, and let us define
\begin{equation}
v(\vect{x},\vect{y},t) = \mathbb{E}\big[\Psi(\vect{x}_t,\vect{y}_t)\,|\,\vect{x},\vect{y}\big],
\end{equation}
that is, the conditional expectation of the observable $\phi(\vect{x},\vect{y})$ as a function of the initial conditions.  
By definition, the function $v(\vect{x},\vect{y},t)$ evolves according to the adjoint of Liouville equation~\eqref{eq:liouville_ms}, i.e.
\begin{equation}
	\partial_t v=-\left(\lambda^2 \mathcal{L}_2 +\lambda \mathcal{L}_1 + \mathcal{L}_0\right)v\,.
	\label{eq:adjoint_liouville}
\end{equation}

To solve the equation above in perturbation theory, we expand $v$ in inverse power of $\lambda$ as
\begin{equation}
v = v^{(0)} + \lambda^{-1} v^{(1)} + \lambda^{-2} v^{(2)} + \cdots=\sum_n \frac{v^{(n)}}{\lambda^n}\,,
\end{equation}
and we plug the expansion into Eq.~\eqref{eq:adjoint_liouville}. The matching of the orders in $\lambda$ yields
\begin{align}
&\mathcal{L}_2 v^{(0)} = 0, \label{eq:ms-0}\\
&\mathcal{L}_2 v^{(1)} = -\mathcal{L}_1 v^{(0)}, \label{eq:ms-1}\\
&\partial_t v^{(0)} + \mathcal{L}_2 v^{(2)} + \mathcal{L}_1 v^{(1)} + \mathcal{L}_0 v^{(0)} = 0\,,\label{eq:ms-2}\\
&\mathcal{L}_2 v^{(n)}+\mathcal{L}_1v^{(n-1)} + \mathcal{L}_0 v^{(n-2)}+\partial_t v^{(n-2)}= 0 \qquad\text{  for  } n\ge3\,.\label{eq:ms-n}
\end{align}
Since $\mathcal{L}_2$ acts only on $\vect{y}$, Eq.~\eqref{eq:ms-0} can be solved by imposing that $v^{(0)}$ is independent on $\vect{y}$, i.e. $v^{(0)} = v^{(0)}(\vect{x},t)$. By considering this solution and by also taking into account the definition of $\mathcal{L}_1$, Eq.~\eqref{eq:ms-1} can be rewritten as
\begin{equation}
\mathcal{L}_2 v^{(1)} = -g(\vect{x},\vect{y}) \cdot \nabla_x v^{(0)},
\end{equation}
whose solution is 
\begin{equation}
v^{(1)} = \Phi(\vect{x},\vect{y}) \cdot \nabla_x v^{(0)},
\end{equation}
where $\Phi$ solves $$\mathcal{L}_2 \Phi = -g\,,$$ 
and the uniqueness of solution $\Phi$ is guaranteed by Fredholm alternative~\cite{pavliotis2008multiscale}.
By plugging the solution for $v^{(1)}$ into Eq.~\ref{eq:ms-2} yields
\begin{equation}
\partial_t v^{(0)} =- \mathcal{L}_2v^{(2)}- \mathcal{L}_1 \left(\Phi(\vect{x},\vect{y}) \cdot \nabla_x v^{(0)}\right) - \mathcal{L}_0 v^{(0)}\,.
\end{equation}
By averaging over the bath probability distribution $\rho_{\vect{x}}$ and considering that, by construction, 
\begin{equation}
    \mathbb{E}_{\vect{x}}\lx[\mathcal{L}_2 v^{(2)}\rx]=\int {\rm d}\vect{y} \rho_{\vect{x}}(\vect{y}) \mathcal{L}_2v^{(2)}=\int {\rm d}\vect{y} \mathcal{L}^*_2\rho_{\vect{x}}(\vect{y}) v^{(2)}=0\,,
\end{equation}
we get a partial differential equation for $v^{(0)}$
\begin{align}
\partial_t v^{(0)} &=- \mathbb{E}_{\vect{x}}\left[\mathcal{L}_1 \left(\Phi(\vect{x},\vect{y}) \cdot \nabla_x v^{(0)}\right)\right] - \mathbb{E}_{\vect{x}}\left[\mathcal{L}_0 v^{(0)}\right]\,,\nonumber\\
 &=- \mathbb{E}_{\vect{x}}\left[\mathcal{L}_1 \left(\Phi(\vect{x},\vect{y}) \cdot \nabla_x v^{(0)}\right)\right] - \mathcal{L}_0 v^{(0)}\,,
\end{align}
where in the second line the expectation value has been dropped since $\mathcal{L}_0 v^{(0)}$ does not depend on $\vect{y}$. 
By using the definition of $\mathcal{L}_1$, the expectation involving this operator reads
\begin{align}
	\mathbb{E}_{\vect{x}}\left[\mathcal{L}_1 \left(\Phi(\vect{x},\vect{y}) \cdot \nabla_x v^{(0)}\right)\right] &= \mathbb{E}_{\vect{x}}\left[g_b(\vect{x},\vect{y}) \cdot \nabla_y \left(\Phi(\vect{x},\vect{y}) \cdot \nabla_x v^{(0)}\right)\right]+\mathbb{E}_{\vect{x}}\left[g(\vect{x},\vect{y}) \cdot \nabla_x \left(\Phi(\vect{x},\vect{y}) \cdot \nabla_x v^{(0)}\right)\right]	\label{eq:averaging}
\end{align}
By denoting with the symbol $C:\nabla_x\nabla_x =\sum_{ij}C_{ij}\partial_{i}\partial_{j}$ the contraction of the gradient with a generic matrix $C$, the second term in the right hand side of Eq.~\eqref{eq:averaging} can we rewritten in the form
\begin{align}
	\mathbb{E}_{\vect{x}}\left[g(\vect{x},\vect{y}) \cdot \nabla_x \left(\Phi(\vect{x},\vect{y}) \cdot \nabla_x v^{(0)}\right)\right]&= \mathbb{E}_{\vect{x}}\lx[g(\vect{x},\vect{y}) \Phi^T(\vect{x},\vect{y})\rx] : \nabla_x \nabla_x v^{(0)}+\mathbb{E}_{\vect{x}}\lx[\left(g(\vect{x},\vect{y}) \cdot \nabla_x \Phi(\vect{x},\vect{y})\right)\rx] \cdot \nabla_x v^{(0)}\label{eq:L1v0_x}
\end{align}

By taking into account the expressions in Eqs.~\eqref{eq:averaging}-\eqref{eq:L1v0_x}, the evolution of a generic observable of the slow degrees of freedom $\vect{x}$ takes the form
\begin{equation}
\partial_t v^{(0)} = -\big[f(\vect{x}) + \Theta(\vect{x})\big]\cdot \nabla_x v^{(0)} - D(\vect{x}):\nabla_x\nabla_x v^{(0)}\,,
\label{eq:adjoint_liouville_slow}
\end{equation}
where the diffusion tensor and drift correction are defined as
\begin{align}
D(\vect{x}) &= \mathbb{E}_{\vect{x}}\lx[g(\vect{x},\vect{y})\Phi^T(\vect{x},\vect{y})\rx]\,, \label{eq:diffusion_ms}\\
\Theta(\vect{x}) &=  \mathbb{E}_{\vect{x}}\left[g_b(\vect{x},\vect{y}) \cdot \nabla_y \Phi(\vect{x},\vect{y}) \right]+\mathbb{E}_{\vect{x}}\lx[\left(g(\vect{x},\vect{y}) \cdot \nabla_x \Phi(\vect{x},\vect{y})\right)\rx]\,. \label{eq:drift_ms-temp}
\end{align}
In principle, once Eq.~\eqref{eq:adjoint_liouville_slow} is solved the correction $v^{(2)}$ should be determined by solving the following differential equation
\begin{equation}
	\mathcal{L}_2v^{(2)} = - \partial_t v^{(0)} - \mathcal{L}_1 \left(\Phi(\vect{x},\vect{y}) \cdot \nabla_x v^{(0)}\right) - \mathcal{L}_0 v^{(0)}
\end{equation}
where the right hand side has to be considered as a source term. The following corrections $v^{(n)}$ for $n\ge3$ are instead determined by the hierarchy~\eqref{eq:ms-n}. Notwithstanding, given that $v^{(0)}$ is an arbitrary function of $\vect{x}$, Eq.~\eqref{eq:adjoint_liouville_slow} suffices to prove that the slow system is well described by a diffusion process in the limit $\lambda\to\infty$. 

\subsection{Derivation of $\Theta(\vect{x})=\Gamma(\vect{x})$}\label{sec:derivation-drift-ms}
To show that under conditions~\eqref{eq:div-free-ms}-\eqref{eq:energy-preservation-ms} the additional drift $\Theta(\vect{x})$ coincides with our definition of $\Gamma(\vect{x})$, we have to explicitly treat the term involving derivative with respect to the bath degrees of freedom $\vect{y}$ appearing in~\eqref{eq:drift_ms-temp}. To this end, let us notice that
\begin{subequations}
\begin{align}
	\mathbb{E}_{\vect{x}}\left[g_b(\vect{x},\vect{y}) \cdot \nabla_y \left(\Phi(\vect{x},\vect{y}) \cdot \nabla_x v^{(0)}\right)\right]&=\int {\rm d}\vect{y}\,\,\rho_{\vect{x}}(\vect{y})g_b(\vect{x},\vect{y}) \cdot \nabla_y \left(\Phi(\vect{x},\vect{y}) \cdot \nabla_x v^{(0)}\right)\label{eq:L1v0_y_a}\\
	&= -\int {\rm d}\vect{y}\,\,\nabla_y\cdot\left(\rho_{\vect{x}}(\vect{y})g_b(\vect{x},\vect{y})\right) \left(\Phi(\vect{x},\vect{y}) \cdot \nabla_x v^{(0)}\right)\label{eq:L1v0_y_b}\\
	&= -\int {\rm d}\vect{y}\,\,\rho_{\vect{x}}(\vect{y})\lx[\frac{\nabla_y\rho_{\vect{x}}}{\rho_{\vect{x}}} \cdot g_b(\vect{x},\vect{y})+\nabla_y\cdot g_b(\vect{x},\vect{y})\rx] \left(\Phi(\vect{x},\vect{y}) \cdot \nabla_x v^{(0)}\right)\label{eq:L1v0_y_c}\\
	&= \int {\rm d}\vect{y}\,\,\rho_{\vect{x}}(\vect{y})\lx[-\frac{\nabla_y\rho_{\vect{x}}}{\rho_{\vect{x}}} \cdot g_b(\vect{x},\vect{y})+\nabla_x\cdot g(\vect{x},\vect{y})\rx] \left(\Phi(\vect{x},\vect{y}) \cdot \nabla_x v^{(0)}\right)\label{eq:L1v0_y_d}\\
	&=\lx(\mathbb{E}_{\vect{x}}\lx[\lx(\nabla_x\cdot g(\vect{x},\vect{y})\rx) \Phi(\vect{x},\vect{y})\rx]-\mathbb{E}_{\vect{x}}\lx[\frac{\nabla_y\rho_{\vect{x}}}{\rho_{\vect{x}}} \cdot g_b(\vect{x},\vect{y})\Phi(\vect{x},\vect{y})\rx]\rx)\cdot \nabla_x v^{(0)}\label{eq:L1v0_y}
\end{align}
\end{subequations}
where from Eq.~\eqref{eq:L1v0_y_a} to Eq.~\eqref{eq:L1v0_y_b} we simply used integration by parts, while from Eq.~\eqref{eq:L1v0_y_c} to Eq.~\eqref{eq:L1v0_y_d} we used the property \eqref{eq:div-free-ms} for expressing the divergence of $g_b$ as a divergence of $g$. 
Let us also notice that 
\begin{equation}
	\mathbb{E}_{\vect{x}}\lx[\lx(\nabla_x\cdot g(\vect{x},\vect{y})\rx) \Phi(\vect{x},\vect{y})\rx]+\mathbb{E}_{\vect{x}}\lx[\left(g(\vect{x},\vect{y}) \cdot \nabla_x \Phi(\vect{x},\vect{y})\right)\rx]=\nabla_x\cdot\mathbb{E}_{\vect{x}}\lx[g(\vect{x},\vect{y})\Phi(\vect{x},\vect{y})^T\rx]-\mathbb{E}_{\vect{x}}\lx[\frac{\nabla_x\rho_{\vect{x}}}{\rho_{\vect{x}}} \cdot g(\vect{x},\vect{y})\Phi(\vect{x},\vect{y})\rx]\,,\label{eq:Theta_terms}
\end{equation}
and thus, by collecting the terms in Eqs.~\eqref{eq:L1v0_y}-\eqref{eq:Theta_terms}, the drift $\Theta(\vect x)$ can be written as
\begin{align}
\Theta(\vect{x}) &=  \mathbb{E}_{\vect{x}}\left[g_b(\vect{x},\vect{y}) \cdot \nabla_y \Phi(\vect{x},\vect{y}) \right]+\mathbb{E}_{\vect{x}}\lx[\left(g(\vect{x},\vect{y}) \cdot \nabla_x \Phi(\vect{x},\vect{y})\right)\rx]\nonumber\\
&=\nabla_x\cdot\mathbb{E}_{\vect{x}}\lx[g(\vect{x},\vect{y})\Phi(\vect{x},\vect{y})^T\rx]-\mathbb{E}_{\vect{x}}\lx[\frac{\nabla_x\rho_{\vect{x}}}{\rho_{\vect{x}}} \cdot g(\vect{x},\vect{y})\Phi(\vect{x},\vect{y})\rx]-\mathbb{E}_{\vect{x}}\lx[\frac{\nabla_y\rho_{\vect{x}}}{\rho_{\vect{x}}} \cdot g_b(\vect{x},\vect{y})\Phi(\vect{x},\vect{y})\rx]\,. \label{eq:drift_ms-temp-2}
\end{align}

%To proceed further we have to consider that the measure of the bath conditioned on the slow variables is a microcanonical distribution
To proceed further we need to evaluate $\nabla_x \log \rho_{\vect{x}}$ and $\nabla_y \log \rho_{\vect{x}}$. Strictly speaking, $\rho_{\vect{x}}(\vect{y})$ is the stationary distribution of the fast flow generated by $f_b(\vect{y})$ at fixed $\vect{x}$. For finite $\lambda$ this differs from the microcanonical distribution of the full system, due to the interaction energy $H_\lambda(\vect{x},\vect{y})$. However, since $H_\lambda = O(r_\lambda) \to 0$, $\rho_{\vect{x}}(\vect{y})$ coincides with the microcanonical distribution at leading order, with corrections of $O(r_\lambda)$. We therefore write at leading order
\begin{equation}
\rho_{\vect{x}}(\vect{y}) = \frac{\delta\left(E-H_{tot}(\vect{x},\vect{y})\right)}{\omega_\lambda(E,\vect{x})}+ O(r_\lambda)\,,
\end{equation}
where
\begin{equation}
	\omega_\lambda(E,\vect{x}) = \int {\rm d} \vect{y}\,\, \delta\left(E - H_{tot}(\vect{x},\vect{y})\right) =  \int {\rm d} \tilde{\beta}\frac{e^{\tilde{\beta}E}}{2\pi i}\int {\rm d} \vect{y}\,\, e^{-\tilde{\beta}H_{tot}(\vect{x},\vect{y})}\,.
 \label{eq:omega_lambda}
\end{equation}
From the Laplace representation of the delta function we have
\begin{align}
&\nabla_x \rho_{\vect{x}}(\vect{y}) = -\frac{1}{\omega_\lambda(E,\vect{x})}\int {\rm d} \tilde{\beta}\frac{e^{\tilde{\beta}E}}{2\pi i}\,\, e^{-\tilde{\beta}H_{tot}(\vect{x},\vect{y})}\lx[\tilde{\beta}\nabla_xH_{tot}+\nabla_x\log\lx(\omega_\lambda\rx)\rx]\,,\\
&\nabla_y \rho_{\vect{x}}(\vect{y}) = -\frac{1}{\omega_\lambda(E,\vect{x})}\int {\rm d} \tilde{\beta}\frac{e^{\tilde{\beta}E}}{2\pi i}\,\, e^{-\tilde{\beta}H_{tot}(\vect{x},\vect{y})}\tilde{\beta}\nabla_yH_{tot}\,,
\end{align}
which, by using Eq.~\eqref{eq:energy-preservation-ms}, imply
\begin{align}
	\mathbb{E}_{\vect{x}}\lx[\lx(\frac{\nabla_y\rho_{\vect{x}}}{\rho_{\vect{x}}} \cdot g_b(\vect{x},\vect{y})+\frac{\nabla_x\rho_{\vect{x}}}{\rho_{\vect{x}}} \cdot g(\vect{x},\vect{y})\rx)\Phi(\vect{x},\vect{y})\rx]&=-\nabla_x\log\lx(\omega_\lambda\rx) \cdot \mathbb{E}_{\vect{x}}\lx[g(\vect{x},\vect{y})\Phi^T(\vect{x},\vect{y})\rx]\,.
\end{align}
Differentiating Eq.~\eqref{eq:omega_lambda} with respect to $\vect{x}$ and using $\nabla_x H_{\mathrm{tot}} = \nabla_x H(\vect{x}) + \nabla_x H_\lambda(\vect{x},\vect{y})$ together with the assumption $\nabla_x H_\lambda = O(r_\lambda)$, we obtain
\begin{align}
	\nabla_x\omega_\lambda(E,\vect{x}) 
    &= -\int {\rm d} \tilde{\beta}\frac{e^{\tilde{\beta}E}}{2\pi i}\int {\rm d} \vect{y}\,\, e^{-\tilde{\beta}H_{\mathrm{tot}}(\vect{x},\vect{y})}\tilde{\beta}\lx[\nabla_xH(\vect{x})+\nabla_xH_{\lambda}(\vect{x},\vect{y})\rx] \nonumber \\
    &= -\nabla_xH(\vect{x})\int {\rm d} \tilde{\beta}\frac{e^{\tilde{\beta}E}}{2\pi i}\,\tilde{\beta}\int {\rm d} \vect{y}\,\, e^{-\tilde{\beta}H_{\mathrm{tot}}(\vect{x},\vect{y})} + O(r_\lambda)\nonumber\\ 
    & = -\nabla_xH(\vect{x})\partial_E \omega_\lambda(E,\vect{x}) +O(r_\lambda)\,.\label{eq:grad_omega}
\end{align}

It remains to compute $\partial_E \log\omega_\lambda$ in the thermodynamic limit $N_\lambda\gg 1$. To this end, let us define the free energy density of the bath
\begin{equation}
    \zeta(\beta) = -\frac{1}{N_\lambda}\log\int {\rm d}\vect{y}\,\, e^{-\beta H_b(\vect{y})}\,,
    \label{eq:free_energy_density}
\end{equation}
such that
\begin{equation}
    \omega_\lambda(E,\vect{x}) = \int \frac{{\rm d}\tilde{\beta}}{2\pi i}\, e^{N_\lambda\lx[\tilde{\beta}\,u - \zeta(\tilde{\beta})\rx]} + O(r_\lambda)\,,
    \label{eq:omega_saddle}
\end{equation}
where 
\begin{equation}
u=\frac{E-H(\vect x)}{N_\lambda}=\frac{E}{N_\lambda}+O\lx(\frac{1}{N_\lambda}\rx).
\end{equation}
 In the limit $N_\lambda\gg 1$, Eq.~\eqref{eq:omega_saddle} is dominated by the saddle point $\beta^\star$ determined by
\begin{equation}
    \lx.\frac{\partial\zeta}{\partial\tilde{\beta}}\rx|_{\beta^\star} = u\,,
    \label{eq:saddle_point}
\end{equation}
which identifies the microcanonical inverse temperature of the bath $\beta^\star=\beta = \partial_E \log\omega_b(E)$. Thus, we arrive at 
\begin{align}
&\omega_\lambda(E,\vect{x}) = \frac{e^{N_\lambda\lx[\beta^\star u - \zeta(\beta^\star)\rx]}}{\sqrt{2\pi N_\lambda\,|\partial_{\beta}^2\zeta(\beta^\star)|}}\lx(1 + O(N_\lambda^{-1})\rx) + O(r_\lambda)\,,\\
&\partial_E \log\omega_\lambda(E,\vect{x})=\frac{1}{N_\lambda}\partial_u \log \omega_\lambda(N_\lambda u,\vect{x})=\beta^*\,.
\end{align}
Finally, the term $\theta(\vect x)$ can be rewritten as
%Thus, we arrive at

%By considering also that
%\begin{equation}
%	\omega_\lambda(E,x) =  \omega_b(E)+O(r_\lambda)\,.
%\end{equation}
%\begin{equation}
%\nabla_x \log\lx(\omega_\lambda\rx)=-\partial_E \log\lx(\omega_b\rx)\nabla_x H = -\beta \nabla_x H
%\end{equation}
%and therefore
%Therefore, 
\begin{align}
    \Theta(\vect{x}) &= \nabla_x\cdot\mathbb{E}_{\vect{x}}\lx[g(\vect{x},\vect{y}) \Phi^T(\vect{x},\vect{y})\rx] -\beta \nabla_x H \cdot \mathbb{E}_{\vect{x}}\lx[g(\vect{x},\vect{y}) \Phi^T(\vect{x},\vect{y})\rx]=\nonumber\\
    &= \nabla_x\cdot D(\vect{x})-\beta D(\vect{x}) \nabla_x H \equiv \Gamma(\vect{x})\,. \label{eq:drift_ms_def}
\end{align}

\subsection{Derivation of time-reversibility condition~(9) for $D(\vect{x})$}\label{sec:derivation-symmetry-diffusion-ms}
Let us now focus on the symmetry of the diffusion tensor $D(\vect{x})$ when the multiscale system~\eqref{eq:det-model-lambda} has the time-reversal symmetry~\eqref{eq:reversibility-ms}. As shown in~\cite{pavliotis2008multiscale}, the diffusion tensor admits the Green--Kubo representation
\begin{align}
D(\vect{x}) &= \int_0^\infty \mathrm{d}s \, \mathbb{E}_{\vect{x}}\lx[g(\vect{x},\vect{y}) g^T(\vect{x},\phi_{\vect{x}}^s(\vect{y}))\rx]\,,\nonumber\\
& = \frac{1}{2}\int_0^\infty \mathrm{d}s \int\mathrm{d}\vect{y} \,\rho_{\vect{x}}(\vect{y})\lx[g(\vect{x},\vect{y}) g^T(\vect{x},\phi_{\vect{x}}^s(\vect{y}))+g(\vect{x},\phi_{\vect{x}}^s(\vect{y})) g^T(\vect{x},\vect{y})\rx]\,,\label{eq:diffusion-kubo}
\end{align}
where the second line make explicit the fact that $D_{ij}(\vect{x})=D_{ji}(\vect{x})$. 
To prove that $D$ as defined in Eq.~\eqref{eq:diffusion-kubo} has the symmetry~(9), we have to evaluate $D(\op (\vect x))$, that is 
\begin{subequations}
\begin{align}
D(\op (\vect{x})) &= \frac{1}{2}\int_0^\infty \mathrm{d}s \int\mathrm{d}\vect{y} \,\rho_{\op (\vect{x})}(\vect{y})\lx[g(\op (\vect{x}),\vect{y}) g^T(\op (\vect{x}),\phi_{\op (\vect{x})}^s(\vect{y}))+g(\op (\vect{x}),\phi_{\op (\vect{x})}^s(\vect{y})) g^T(\op (\vect{x}),\vect{y})\rx]\,,\label{eq:diffusion-kubo-eps-a}\\
&= \frac{1}{2}\int_0^\infty \mathrm{d}s \int\mathrm{d}\op_b \lx(\vect{z}\rx) \,\rho_{\op (\vect{x})}(\op_b (\vect{z}))\lx[g(\op (\vect{x}),\op_b (\vect{z})) g^T(\op (\vect{x}),\phi_{\op (\vect{x})}^s(\op_b (\vect{z})))+g(\op( \vect x),\phi_{\op (\vect{x})}^s(\op_b (\vect{z}))) g^T(\op (\vect{x}),\op_b (\vect{z}))\rx]\label{eq:diffusion-kubo-eps-b}\\
&= \frac{1}{2}\int_0^\infty \mathrm{d}s \int\mathrm{d}\vect{z}\,\rho_{\vect{x}}(\vect{z})\lx[g(\op (\vect{x}),\op_b (\vect{z})) g^T(\op (\vect{x}),\op_b (\phi_{ \vect{x}}^{-s}( \vect{z})))+g(\op (\vect{x}),\op_b (\phi_{ \vect{x}}^{-s}(\vect{z}))) g^T(\op (\vect{x}),\op_b (\vect{z}))\rx]\label{eq:diffusion-kubo-eps-c}\\
&= \frac{1}{2}\int_0^\infty \mathrm{d}s \int\mathrm{d}\vect{z}\,\rho_{\vect{x}}(\vect{z})\lx[M(\vect x)g(\vect{x},\vect{z}) g^T(\vect{x}, \phi_{ \vect{x}}^{-s}( \vect{z}))M^T(\vect x)+M(\vect x)g( \vect{x}, \phi_{ \vect{x}}^{-s}(\vect{z})) g^T( \vect{x}, \vect{z})M^T(\vect x)\rx]\label{eq:diffusion-kubo-eps-d}\\
&= M(\vect x)\lx(\frac{1}{2}\int_0^\infty \mathrm{d}s \int\mathrm{d}\vect{y}\,\rho_{\vect{x}}(\vect{y})\lx[g(\vect{x},\phi_{ \vect{x}}^{s}( \vect{y})) g^T(\vect{x},\vect{y})+g( \vect{x}, \vect{y}) g^T( \vect{x}, \phi_{ \vect{x}}^{s}( \vect{y}))\rx]\rx)M^T(\vect x)\label{eq:diffusion-kubo-eps-e}\\
&= M(\vect x)D(\vect{x})M^T(\vect x)\label{eq:diffusion-kubo-eps-f}
\end{align}
\end{subequations}
where from Eq.~\eqref{eq:diffusion-kubo-eps-a} to Eq.~\eqref{eq:diffusion-kubo-eps-b} we performed the change of variable $\vect y =\eps_b(\vect{z})$ and we used $|\det{M_b}|=1$, from Eq.~\eqref{eq:diffusion-kubo-eps-b} to Eq.~\eqref{eq:diffusion-kubo-eps-c} we considered the properties of the flux $\phi_{\op (\vect x)}^s( \vect y)=\phi_{\vect x}^{s}(\vect y)$ and $\phi_{\vect x}^s( \op_b (\vect y))=\op_b (\phi_{\vect{x}}^{-s}(\vect y))$. From Eq.~\eqref{eq:diffusion-kubo-eps-c} to Eq.~\eqref{eq:diffusion-kubo-eps-d} we used the property og $g$ under time-reversal $\op_b$ while from Eq.~\eqref{eq:diffusion-kubo-eps-d} to Eq.~\eqref{eq:diffusion-kubo-eps-e} we performed the change of variable $\vect y =\phi_{\vect x}^{-s}(\vect z)$ and used the stationarity of the measure ${\rm d}\vect{z}\,\rho_{\vect{x}}(\vect z)$.

\section{Examples}
\label{sec:example}
In this Section we provide a series of examples of dynamics characterized by non-trivial symmetries under time reversal, where $\op$ is neither the identity nor the familiar $(\vect q,\vect p)\to(\vect q,-\vect p)$ tranformation.

\subsection{Two-dimensional hydrodynamics}
An example is represented by point vortexes in two-dimensional hydrodynamics: denoting by $\vect{r}_j=(X_j,Y_j)$, $1 \le j \le N$, their positions, the Hamiltonian reads
\begin{equation}
\label{eq:vort}
H(\vect{X},\vect{Y})=-\frac{1}{2 \pi}\sum_{j>k}c_j c_k \log|\vect{r}_j-\vect{r}_k|\,,
\end{equation}
where $\{c_j\}_{j=1}^N$ are constants. For the sake of simplicity, we consider the infinite plane as the domain of the evolution, but similar results hold for other confinements. Since $H(\vect{X},\vect{Y})=H(\vect{Y}, \vect{X})$, it can be shown that the operator $\op$ that switches each $X_j$ and $Y_j$ provides the desired reversing rule. Equilibrium stochastic descriptions corresponding to model~\eqref{eq:vort} have been studied in the literature~\cite{chavanis2001kinetic}: they turn out to have form~(4) and to fulfill Eqs.~(7) and~(9).

\subsection{The truncated Euler equation}
As noted in~\cite{bouchet2014langevin,bouchet2014non}, the relation between equilibrium stochastic processes and corresponding deterministic dynamics does not require an Hamiltonian structure. Examples are provided by a class of dynamical systems relevant for equilibrium hydrodynamics. %but the method we present can be easily extended to all deterministic systems having .
Let $\vect{x}\in \mathbb{R}^n$ be the state of the system with evolution
\begin{equation}
\dot{x}_k = \sum_{ml}A_{kml}x_m x_l\,.
\end{equation}
where the values of $A_{kml}$ ensure a conservation law of the form
$H(\vect{x})=\frac{1}{2}\sum_k x^2_k= E$. 
Interpreting the $x_k$ as the Fourier modes coefficients of the velocity $u$ of a fluid, this system represents the truncated Euler equations~\cite{kraichnan1980two,majda2001mathematical}. 
It can be easily verified that the transformation $\vect{x}\to-\vect{x}$ is a symmetry of the energy $H$ that reverses the dynamics~\cite{bouchet2014langevin,bouchet2014non}. Therefore, the general equilibrium stochastic process corresponding to such system reads
\begin{equation}
\dot{x}_k = \sum_{ml}A_{kml}x_m x_l- \gamma \beta\sum_l D_{nl}(\vect{x})x_l+\xi_n\,.
\end{equation}
where $\av{\xi_k(t)\xi_m(t')}=\gamma D_{km}(\vect{x})\delta(t-t')$ and the inverse temperature $\beta$ depends on the energy $E$. Note that the above system reproduces a Gaussian statistics as expected in inviscid equilibrium hydrodynamics~\cite{boffetta2012two}. Moreover, focusing on a diagonal noise matrix $D= I_n/\beta$, the dynamics reduces to
\begin{equation}
\dot{x}_k = \sum_{ml}A_{kml}x_m x_l- \gamma x_k+\xi_k\,.
\end{equation}
which is considered appropriate for describing the large scale behaviour of turbulent flows~\cite{majda2001mathematical}, especially in the two-dimensional case~\cite{bouchet2014langevin,bouchet2014non,herbert2012statistical}.

\subsection{The asymmetric Lotka-Volterra model}
The Lotka-Volterra model~(21) can be written as a Hamiltonian system with
\begin{equation}
    \label{eq:lvham}
    H(q,p)=e^p-p-1 + \frac{a}{c}\cbr{e^q-q-1}\,
\end{equation}
where $q=\log (b y/a)$, $p= \log (d x/c)$
and time is rescaled by $c^{-1}$.
If $a\ne c$ the system does not exhibit any simple symmetry, and the time-reversal rule $\op$ is not trivial.

\begin{figure*}
    \centering    \includegraphics[width=0.99\linewidth]{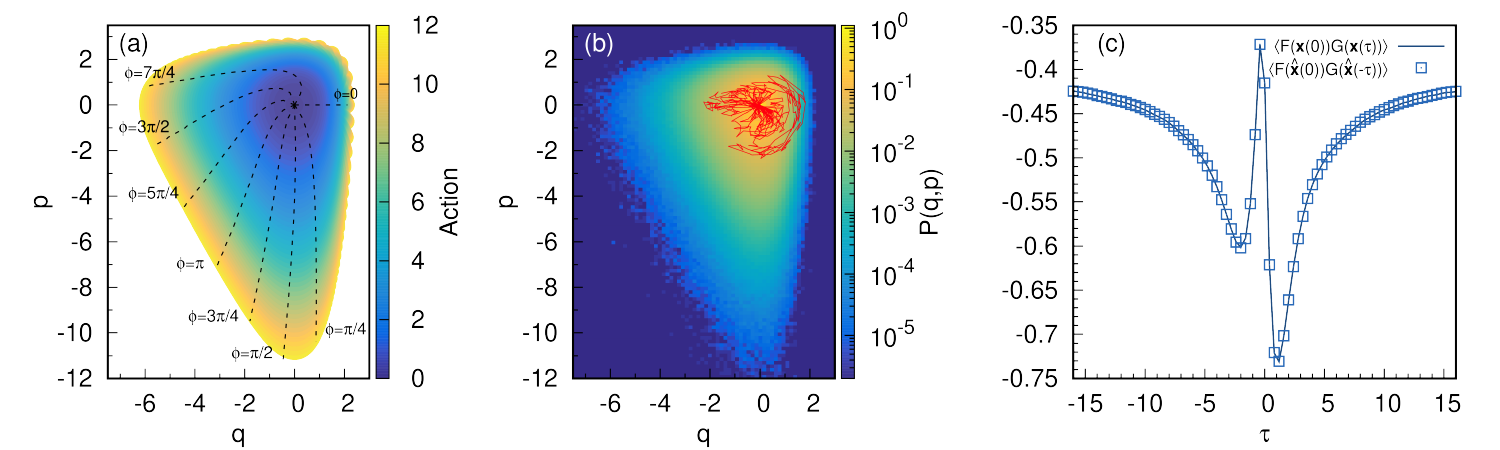}
    \caption{Time-reversal symmetry in the Lotka-Volterra model. In panel (a), $p,q$ coordinates are mapped into the corresponding action $I$ (color scale) and angle $\phi$ (multiples of $\pi/4$ are marked by dashed lines). Panel (b) represents instead a typical trajectory of the stochastic system~\eqref{eq:lvasaa} (red curve, integrated over 200 time units), compared with the empirical distribution of the system in $(q,p)$ space (color scale). Panel (c): test of detailed balance through Eq.~(23), with $F(q,p)=p$ and $G(q,p)=q^2$. The l.h.s. (solid line) and r.h.s. (squares) of Eq.~(23) are compared, showing perfect agreement.  We used a Euler-Maruyama integration algorithm with time-step $\Delta t=10^{-3}$ and total simulation time $t_{max}=10^7$. Parameters: $a/c=2$, $\beta=1$.}
    \label{fig:lvasymm}
\end{figure*}

Our discussion in the main text provides a straightforward method to define 2D reversible stochastic processes with a given deterministic limit~\eqref{eq:dyn}, with $f(\vect{x})=J \nabla H(q,p)$, on the same lines as~\cite{bouchet2014langevin,bouchet2014non,dubkov2009non}. Indeed, the Langevin equation (in action-angle variables)
\begin{subequations}
    \label{eq:langevinaa}    
\begin{eqnarray}
    \dot{\phi}&=&\partial_I K+\gamma\partial_{\phi}\widetilde{D}_{\phi \phi}+\gamma\partial_I \widetilde{D}_{\phi I}-\gamma\beta \widetilde{D}_{\phi I} \partial_I K +\xi_{\phi}\\
    \dot{I}&=&\gamma\partial_{\phi}\widetilde{D}_{\phi I}+\gamma\partial_I \widetilde{D}_{I I}-\gamma\beta \widetilde{D}_{I I} \partial_I K +\xi_{I}\,,
\end{eqnarray}
\end{subequations}
with $\av{\vect{\xi}(t)\vect{\xi}^T(t')}=2\gamma \widetilde{D} \delta(t-t')$, satisfies the required property if $\widetilde{D}=\begin{pmatrix}\widetilde{D}_{\phi \phi} & \widetilde{D}_{\phi I} \\ \widetilde{D}_{\phi I} & \widetilde{D}_{I I} \end{pmatrix}$ fulfills $\widetilde{M}(\vect{x})\widetilde{D}(\vect{x})\widetilde{M}^T(\vect{x})=\widetilde{D}(\op(\vect{x}))$. Here $\widetilde{M}$ is computed from the TR operator $\widetilde{\op}$ defined by Eq.~(18).
The $(q,p)$ description is then recovered by applying $\mathcal{W}$.
In Fig.~\ref{fig:lvasymm}, we show numerical results for a stochastic dynamics of the form
\begin{subequations}
\label{eq:lvasaa}
\begin{eqnarray}
    \dot{I}&=&-\beta D_{II}\partial_I K(I)+\xi_I\\
    \dot{\phi}&=&\partial_I K(I)+\xi_{\phi}\,,
\end{eqnarray}
\end{subequations}
where $K(I)$ is the Hamiltonian of the original deterministic system, $\xi_I$ and $\xi_{\phi}$ are independent Gaussian white noises with variance $2 D_{II}$ and $2D_{\phi \phi}$, respectively, with $D_{II}$, $D_{\phi \phi}$ independent of $I$ and $\phi$. This process is additive and it can be simulated via a Euler-Maruyama integration scheme. When the value of $\phi$ exceeds the domain $[0,2\pi)$, periodic boundary conditions are applied. This is consistent with the physical interpretation of $\phi$ as the phase of the motion along a closed orbit. When $I$ becomes negative, we apply instead the boundary condition  $(0^{-},\phi)\equiv(0^{+},2\pi-\phi)$: in $(q,p)$ representation, this can be seen as the trajectory crossing the origin. Other choices would be admissible. For instance, one may choose the matrix $D$ in such a way that $I$ never vanishes. The price to pay would be the presence of multiplicative noise in the dynamics.
To come back to the original $(q,p)$ variables, we proceed as follows. For a set of equally spaced values $I_n\in(0,I_{max}]$, $n=1,...,N$, one computes $E_n=K(I_n)$. Then, for each of these energies, the phase space point $(q^{(n)}_0,p^{(n)}_0)\in \mathcal{C}(E_n)$, with $p^{(n)}_0=0$ and $q^{(n)}_0\ge 0$, is set as the one corresponding to $\phi=0$. $M$ pairs $(q^{(n)}_m,p^{(n)}_m)$, relative to $\phi_m=2\pi m/M$, $m=1,...,M$ are found by evolving $(q^{(n)}_0,p^{(n)}_0)$ by a time $t_m=m T(E_n)/M$ (a task that can be done numerically by using a symplectic algorithm, such as the Verlet update). In this way, a discrete map
$$
(I_n, \phi_m) \to (q^{(n)}_m,p^{(n)}_m)
$$
is built, which can be interpolated to provide the desired $(q,p)\leftrightarrow(I,\phi)$ relation also in the continuous domain.

%In order to build a reversible stochastic process with equilibrium stationary distribution $\propto \exp[-\beta H(q,p)]$, we pass to action-angle variables $I(q,p)$, $\phi(q,p)$, as discussed in the main text. 
The mapping between the two descriptions is illustrated in Fig.~\ref{fig:lvasymm}(a) for the case $a/c=2$. 
Figure~\ref{fig:lvasymm}(b) shows a typical trajectory of process~\eqref{eq:lvasaa}. By construction, the stationary pdf is the Boltzmann distribution. We also verify in Fig.~\ref{fig:lvasymm}(c) that detailed balance holds, under the time-reversal operator~(18) (with $\phi_0(I)=2\pi$), exploiting relation~(23).

\begin{figure}
    \centering
    \includegraphics[width=0.99\linewidth]{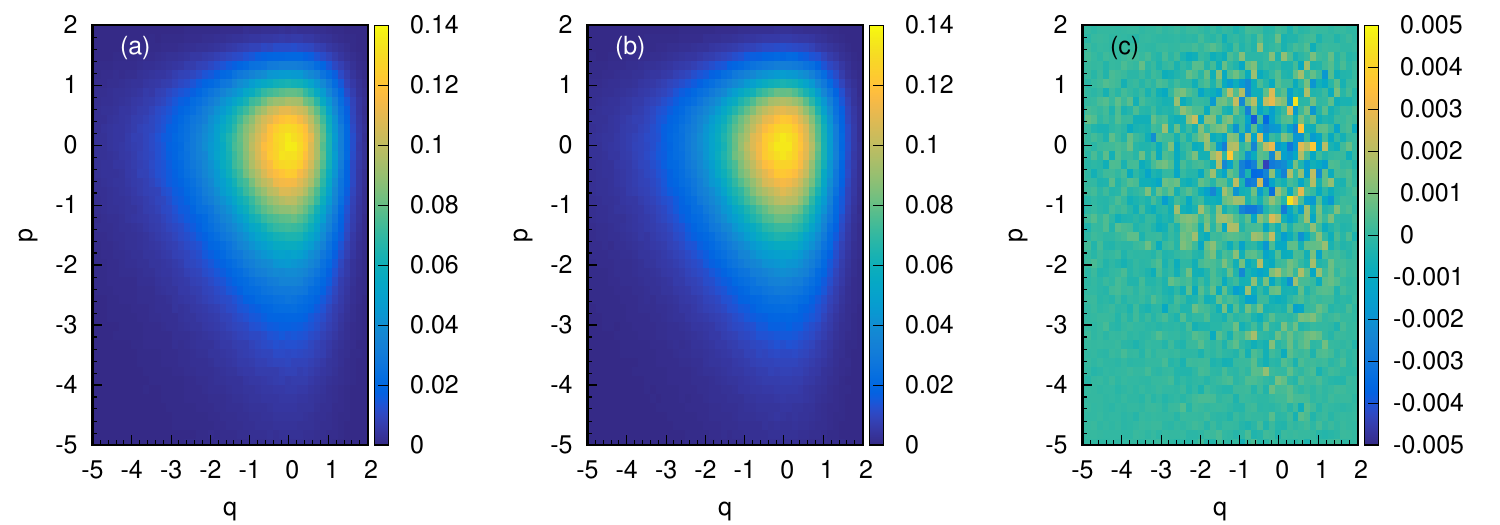}
    \caption{Empirical stationary distributions of the stochastic process described in Fig.~2 of the main text. Panels (a) and (b) of this figure represent the empirical pdf for the equilibrium and out-of-equilibrium dynamics [cases (a) and (c) in Fig.~2 of the main text, respectively]. Panel (c) shows their difference, bin by bin. The bin size is $0.15\times0.15$. Each histogram has been obtained sampling $10^7$ points at regular intervals along a trajectory of total length $t_{max}=10^6$. Other parameters as in Fig.~2 of the main text.}
    \label{fig:distributions}
\end{figure}

Consistently with our derivation, the empirical distributions of the equilibrium and out-of-equilibrium process are the same. We report them in Fig.~\ref{fig:distributions}, as well as their bin-by-bin difference, showing that they are fully equivalent. Both pdf's are compatible with a Boltzmann distribution with $\beta=1$, as expected.

\subsection{Nonequilibrium stochastic formulation of Lotka-Volterra model}
%    Let us consider the Lotka-Volterra model for population dynamics 
Generalized Lotka-Volterra models are widely used in the theoretical study of ecologic systems~\cite{goel1971volterra,altieri2021properties,suweis2024generalized}. An extensive statical-mechanics analysis of this model is provided in~\cite{goel1971volterra}, where the authors also provide a stochastic formulation of the evolution. Here, we want to show that such stochastic dynamics is intrinsically out of equilibrium. For the sake of simplicity, we only focus on the case $a=c$, although the same reasoning can be applied to the general case. In the present case, the Hamiltonian~\eqref{eq:lvham}
is symmetric under $\op(q,p)=(p,q)$, which is also a reversing symmetry for the deterministic evolution
 \begin{equation}
 \begin{cases}
    \dot{q}=( e^p-1)\\
    \dot{p}=-( e^q-1)\,.
\end{cases}
\end{equation}
A stochastic version of this model has been derived by allowing the coupling between the two populations to fluctuate in time~\cite{goel1971volterra}. The resulting Langevin dynamics is
 \begin{equation}
 \label{eq:lv-montroll}
\begin{cases}
    \dot{q}=( e^p-1)+\xi_q\\
    \dot{p}=-( e^q-1)+\xi_p\,.
\end{cases}
 \end{equation}
where $\av{\boldsymbol{\xi}(t)\boldsymbol{\xi}^T(t')}=2D(q,p)\delta(t-t')$, with
\begin{equation}
D(q,p)=\sigma^2\begin{pmatrix}
e^{2p} & -e^{p+q}\\
-e^{p+q} & e^{2q}
\end{pmatrix}\,,
\end{equation}
where $\sigma$ is constant. Note that $D$ satisfies relation $M(\vect{x})D(\vect{x})M^T(\vect{x})=D(\op(\vect{x}))$. However, since a  dissipation
$\Gamma$ compatible with~\eqref{eq:dissipation_final}
is not included in Eq.~\eqref{eq:lv-montroll}, the stationary solution of this process does not coincide with $e^{-\beta H}$. More importantly, we can prove that such dynamics does not fulfill detailed balance. 
As shown in Section~\ref{sec:cond}, the general formulation of detailed balance, without assumptions on $\cP_S$, reads
\begin{equation}
\label{eq:detailed_balance_lv}
 \sum_k M^{-1}_{jk} A_k (\op (\vect{x}))=-A_j(\vect{x})+2\cP_S^{-1}(\vect{x})\sum_l \partial_l D_{jl}(\vect{x})\cP_S(\vect{x})
\end{equation}
with $A ( \vect{x})=J\nabla H$.  Since $\op(\vect{x})=M\vect{x}$ is a reversing symmetry of the deterministic dynamics, the drift vector satisfies $A (\op  \vect{x})=-M A(\vect{x})$,  and Eq.~\eqref{eq:detailed_balance_lv} reduces to
\begin{equation}
\label{eq:detailed_balance_montroll}
 0=2\cP_S^{-1}(\vect{x})\sum_l \partial_l D_{jl}(\vect{x})\cP_S(\vect{x})
\end{equation}
Inserting the above equation into the Fokker-Planck evolution leads to
\begin{equation}
\label{eq:fp_lv}
    \sum_i \partial_i \lx[A_i(\vect{x})\cP_S(\vect{x})  \rx]=0
\end{equation}
whose solution is $\cP_S(\vect{x})=f(H(\vect{x}))$. An explicit check shows that it does not exist an $f$ which simultaneously satisfies both Eq.~\eqref{eq:fp_lv} and Eq.~\eqref{eq:detailed_balance_montroll}, and, therefore, Eq.\eqref{eq:detailed_balance_lv} does not hold.

\bibliography{biblio}